# 100% renewable electricity in Japan

Cheng Cheng, Andrew Blakers, Matthew Stocks, Bin Lu

RE100 Group, School of Engineering

Australian National University, Canberra, Australia

## Abstract

Japan has committed to carbon neutrality by 2050. Emissions from the electricity sector amount to 42% of the total. Solar photovoltaics (PV) and wind comprise three quarters of global net capacity additions because of low and falling prices. This provides an opportunity for Japan to make large reductions in emissions while also reducing its dependence on energy imports. This study shows that Japan has 14 times more solar and offshore wind resources than needed to supply 100% renewable electricity. A 40-year hourly energy balance model is presented of Japan's electricity system using historical data. Pumped hydro energy storage, high voltage interconnection and dispatchable capacity (hydro, biomass and hydrogen energy) are included to balance variable generation and demand. Differential evolution is used to find the least-cost solution under various constraints. The levelized cost of electricity is found to be US$86/MWh for a PV-dominated system, and US$110/MWh for a wind-dominated system. These costs can be compared with the average system prices on the spot market in Japan of US$102/MWh. In summary, Japan can be self-sufficient for electricity supply at competitive costs.

1. Introduction

*1.1 Policy context*

Japan is the fifth largest greenhouse gas (GHG) emitter in the world. Japan currently generates 21% of its electricity from renewables, with the balance comprising nuclear (7%), fossil fuels (70%) and other (2%)[1]. The decision of the Japanese Government to commit to zero emissions in 2050[2] means that large-scale decarbonization of energy needs to take place in the following decades. The electricity system, accounting for 42% of the total emissions in 2019[3], is the best place to start, since decarbonized electricity can also displace oil in the land transport sector (electric vehicles) and gas in the heating sector (with assistance from heat pumps and electric furnaces).

Following Japan's prime ministerial pledge to achieve carbon neutrality by 2050, Japan's Ministry of Economy, Trade and Industry (METI) formulated a Green Growth Strategy[3] to provide a policy framework for this transition. According to this Strategy, Japan will generate 50% - 60% of its electricity from renewables by 2050, mainly from offshore wind. The rest is expected to be supplied by hydrogen (10%) and nuclear and fossil fuel power plants with carbon capture and storage (CCS) (30% - 40%). In the report METI argues that 100% renewable electricity is difficult due to environmental and social constraints and concerns regarding energy flexibility, transmission, system inertia and costs. However, many studies, including those presented in a recent meeting held within METI's Comprehensive Resources and Energy Study Group Basic Policy Subcommittee[4], have shown that 100% renewable electricity is technically feasible at competitive costs for many countries and regions[5].

The cost of solar photovoltaic (PV) and wind energy is falling and, in many places, is cheaper than the cost of electricity from new-build coal and gas power stations[6]. Solar PV and wind now account for three quarters of global net capacity additions due to their low and falling prices[7]. Wide deployment of solar PV and wind globally means that global cost convergence is likely to happen in the next few decades as more experience is gained and local markets become more competitive. Consequently, future costs of solar PV and wind in Japan are expected to be much lower than today's level.

As will be shown in this paper, solar PV and offshore wind are the most promising ways to decarbonize electricity in Japan. Off-river pumped hydro energy storage (PHES) and transmission allow variable generation to be balanced at low cost using mature and reliable technology with abundant availability and low environmental impact[8]. Due to safety concerns after the Fukushima accident, a consensus has been reached in Japan that dependence on nuclear needs to be minimized[9]. CCS is still in its demonstration phase, with only 26 operating facilities globally with an annual capacity of 40 million tonnes of $CO_2$[10]. It is far from ready for mass-deployment in the required timeframe to reach zero emissions in 2050 and future costs of CCS are unknown as these early-stage projects provide little information.

Japan could rely largely on imported zero-emissions energy-rich chemicals, such as ammonia produced using hydrogen derived from electrolysis powered by wind and solar in another country. However, the cost is high because the round-trip efficiency is low (~25%[11]) in converting renewable electricity to hydrogen, followed by shipment to Japan, and finally re-conversion to electricity. Green hydrogen may be beneficial when utilized occasionally to provide peaking power, but significant cost reductions would be needed for it to contribute large fractions. To this end, decarbonizing electricity using renewable energy avoids the safety issues associated with nuclear, and effectively eliminates the need to rely on future developments and cost reductions of CCS and hydrogen technology. Relying on domestic renewable energy resources such as solar and wind also allows Japan to reduce dependence on energy imports, considering that Japan lacks fossil fuel reserves and currently imports most of its fossil and nuclear fuels[9].

This study focuses on the future electricity system in Japan and addresses the following questions:

1) Does Japan have sufficient renewable energy resources to supply its entire electricity demand?
2) What is the levelized cost of electricity (LCOE) of the proposed 100% renewable electricity system?

The outcomes of this study are expected to inform policymakers of the potential of domestic renewable energy resources and an alternative pathway to decarbonized electricity in Japan. This study also provides information for industry participants and international partners and may affect investment decisions.

*1.2 Balancing an electricity system dominated by intermittent renewables*

Continued cost reductions and increased uptake of solar PV and wind globally have led to deliberations on the costs and feasibility of 100% renewable electricity systems in the literature in recent years[5]. Storage, strong interconnection between regions and demand management help balance supply and demand to support large proportions of variable renewable generation[7]. Energy storage from electricity include chemical (e.g., hydrogen or batteries), thermal (molten salts), kinetic (flywheels) and potential energy (pumped hydro).  Pumped hydro energy storage (PHES) constitutes more than 95% of global storage energy volume and storage power for the electricity industry. Pumped hydro is the lowest cost, most mature and largest-scale storage technology and is capable of supporting 100% renewable electricity systems at low cost[12,13]. It can also provide ancillary services for the grid including mechanical inertia in place of retiring coal and gas power plants. Most existing pumped hydro storage is associated with river-based hydroelectricity systems, which can have large environmental cost and strong social pushback. However, most of the world's land is not near a river, and there are many "off-river" sites. A global atlas of off-river pumped hydro energy storage identified 616,000 promising sites with combined storage of 23 million Gigawatt-hours (GWh) (an enormous amount of storage) distributed across most regions of the world[14], including 2,400 sites in Japan with a combined storage potential of 53,000 GWh.

Japan had 28 Gigawatts (GW) of existing pumped hydro energy storage installed as of 2018[9], most of which is river-based and was built prior to the 2011 Fukushima disaster to balance generation from nuclear plants. The existing pumped hydro schemes in Japan are useful for balancing intermittent generation from solar PV and wind in a 100% renewable grid. With continued cost reductions, batteries will also play an important role in short-term supply-demand balancing and in electric vehicles. However, pumped hydro is cheaper than batteries for overnight storage[15], which is required in a system with large fractions of solar PV. Pumped hydro can provide large-scale energy storage while batteries are well suited to provision of storage power needed for ancillary services.

*1.3 100% renewable electricity in Japan*

While 100% renewable electricity systems in many countries and regions have been largely discussed in the literature[5], only a few studies from academia have investigated the future role of renewable energy from Japan's perspective[16–20]. There are major gaps in the literature as follows:

1. Some studies (e.g.[16,20]) included coal or fossil gas in the generation mix but did not include corresponding technologies (e.g. CCS) to deal with the emissions from these fossil fuel plants. Therefore, these studies did not provide effective solutions to carbon neutrality.

2. Most studies did not analyze such systems from an economic perspective and therefore did not derive least-cost solutions[17–20].
3. Most studies assumed unlimited PV and wind potential and did not take the resource constraint into account[17–20]. In fact, Japan has limited land area, and it is essential that the required generation and storage capacities can be deployed at low cost in Japan. On this account, Japan has recently started to explore its offshore wind resources, with four offshore wind promotion zones announced by the Japanese government in 2020[21]. Offshore wind resources in Japan are overlooked in many studies[16,17,19].
4. All studies overlooked the vast potential of off-river pumped hydro energy storage to provide mature and low-cost storage for hours to weeks[16–20]. Instead, these studies only used existing pumped hydro capacity and expensive batteries provided additional primary storage.
5. All studies used meteorological data from a single year[16–19] or several years[20] for modelling. However, Japan periodically experiences extreme weather[22] and models that are based on meteorological data over a short period provide little information on the long term system performance.

In addition to the studies conducted by academics, in a recent meeting held within METI's Comprehensive Resources and Energy Study Group Basic Policy Subcommittee, several decarbonization scenario analysis pieces were presented by local organizations[4], including Renewable Energy Institute (REI), The Institute of Energy Economics Japan (IEEJ), Deloitte Tohmatsu Consulting and The Research Institute of Innovative Technology for the Earth (RITE) etc. A summary of the 100% renewable studies presented in this meeting is shown in the Table below.

*Table 1: Summary of 100% renewable energy scenario analysis presented in the METI meeting. Information sourced from METI website[4].*

| Organization | RITE | REI | Deloitte | IEEJ |
|---|---|---|---|---|
| **Model objective** | Least-cost | Least-cost | Least-cost | Least-cost |
| **Electricity demand** | 1100 TWh | 1470 TWh | 1450 TWh | 1200 TWh |
| **Generation mix** | 100% renewables (Detailed installed capacity not provided) | 100% renewables, including 8% import:<br>• 524 GW solar PV<br>• 63 GW offshore wind<br>• 88 GW onshore wind<br>• 22 GW hydropower<br>• 6 GW bio & geothermal | • 95% renewables<br>• 2% nuclear<br>• 3% CCS<br>(Detailed installed capacity not provided) | 100% renewables (Detailed installed capacity not provided) |

|  |  | • 20 GW import |  |  |
|---|---|---|---|---|
| **Storage** | • 3980 GWh batteries<br>• 570 GW system enhancement | • 42GW/178GWh utility battery<br>• 45GW/276GWh household battery<br>• 30GW/180GWh V2G<br>• 30GW/180GWh PHES<br>• 82 GW interconnection<br>• 20 GW international connection<br>• 73 GW electrolyser | • 44GWh battery<br>• 945GW system enhancement<br>• 129 TWh V2G | • 398GWh battery<br>• 3434GWh compressed hydrogen storage<br>• 468GWh V2G<br>• 64.4GW system enhancement |
| **LCOE ($US)** | $165/MWh | $84/MWh | $174/MWh | $247/MWh |

These studies presented various pathways to energy decarbonization in Japan. However, as pointed out by METI, the pathways presented will be reviewed flexibly in the coming decades based on the progress of cost reduction and technology development, and Japan is not planning to specify a certain energy source composition in the short term, but instead is more interested in having a range of policy options available.

It is also important to point out that a wide range of values were quoted as 'upper limits' for solar PV and wind capacities in these studies. The estimated solar PV potential in Japan ranged between 350 GW to 2746 GW among multiple studies, while that for wind ranged between 296 GW to 938 GW[4]. Most of these studies overlooked alternative types of solar PV, such as agrivoltaics (solar arrays installed on top of crops) and floating PV (solar arrays on water bodies), despite their rapid deployment globally[23,24]. In addition, offshore wind resources at deep water (>200m) is often excluded, while projects are already being planned at locations with greater water depth[25,26], and projects with water depth up to 1000m is technically feasible[27–29]. It is, therefore, a need for a detailed resource assessment for Japan that considers all potential sites for future solar PV and wind deployment to avoid over-conservative assumptions.

In this study, we present a long-term (40 years), high-resolution (hourly) energy balancing analysis of Japan's future electricity system supplied by 100% renewable electricity, mainly solar PV and offshore wind. Hourly balancing of intermittent supply and demand is provided by pumped hydro energy storage, high voltage direct current (HVDC) and alternating current (HVAC) transmission and a small portion of other flexible energy resources (existing hydro, bio energy and green hydrogen).

Using Differential Evolution[30] we derive the least-cost electricity system configuration under defined reliability, resource, energy and transmission constraints for multiple scenarios. Batteries are excluded from the scope of this study due to the current high costs. However, with future cost reductions, batteries may become cheaper than pumped hydro for short-term storage, leading to lower balancing costs than the results presented. Thus, costs estimated in this study are effectively an upper bound. Onshore wind is also excluded in this study because it is not involved in Japan's current plan for carbon neutrality[3] due to lower capacity factors compared with offshore wind[31] and limited number of available sites in Japan[32].

This study is an important addition to this group of 100% renewable energy studies for Japan. It introduces an alternative pathway for Japan that is built upon off-river pumped hydro for low-cost, large-scale, mature energy storage, which is yet to be well-examined in academic, business or political reports. It also presents a wide range of energy independence scenarios in which all electricity is supplied and balanced by domestic renewable energy resources, thereby avoiding energy security issues associated with energy imports.

Although this study models only historical electricity demand (around 900 TWh p.a.) and does not include electrification of transport, heat, and industry, a comprehensive Geographic Information System (GIS) based resource assessment is presented to identify the technical resource potential of solar PV and offshore wind in Japan. As will be shown later in this study, domestic renewable energy resources in Japan are far larger than needed and the proposed pathway is well-suited to incorporate additional electricity demand due to electrification of other energy sectors. Detailed modelling of other energy sectors will be included in future work.

To the best of the authors' knowledge, this is the first time that the future role and costs of a PV-Offshore Wind-Pumped-Hydro hybrid system in supplying 100% renewable electricity in Japan is discussed in detail. The optimized system configurations and costs in various scenarios can be used as references by policymakers and grid operators when making decisions for the transition to a carbon neutral society.

2. Methods

The methods of this study are introduced in this section. A brief description of the data used in this study is available in **Supplementary Information A**.

*2.1 Optimization process to derive the least-cost solution*

This study uses a modified version of the modelling framework introduced by Lu et al.[13]. The model uses time series demand and meteorological data to simulate the hourly energy balance, including demand, generation and charging/discharging of storage in each service area. The model aims at deriving the least-cost electricity system configuration under the following constraints:

- Reliability constraint: electricity generation must meet demand in every timestep unless a specified amount of deficit is allowed, to represent load shedding in certain scenarios.
- Resource constraint: installed capacity of a technology in a service area must not exceed the identified technical resource potential of this technology in this service area.
- Energy constraint: total generation from a certain technology must not exceed the specified maximum generation from this technology.
- Transmission constraint: power flow in a transmission line must not exceed the specified maximum capacity of this transmission line.

For a given set of optimization parameters (e.g. PV, wind and storage capacity in each service area) with defined upper bounds determined by the resource constraint, the model uses Differential Evolution[33] to find the parameters with which an objective function returns the lowest value. The objective function consists of the LCOE calculated based on energy balance simulations, and the penalties for not meeting the reliability, energy, or transmission constraints.

In this study, the model is modified to incorporate hydrogen as an additional dispatchable source, to reduce the amount of pumped hydro storage required to cover occasional periods of low solar and wind availability. Hydrogen can be either imported (in the short term) or derived from water via electrolysis driven by curtailed solar and wind generation in Japan. Detailed information on the algorithm and modifications of the model is available in the **Supplementary Information B**.

*2.2 GIS-based resource assessment for solar PV and offshore wind*

In order to set the resource constraints (upper bounds for the optimization parameters) in the model, the maximum capacity of PV, wind and storage in each service area need to be determined. GIS analysis is performed to estimate the technical resource potential of PV and wind in each region, while data from the global atlas of off-river pumped hydro energy storage[14] is used to set the limit of pumped hydro capacity.

Four forms of solar PV deployments are considered in Japan: ground-mounted PV (GPV), building-integrated PV (BIPV), floating PV (FPV) (on rivers, reservoirs, and the inland sea) and agrivoltaics (APV) (solar array installed above crops). By incorporating land use data[34], protected area data[35], wave height data[36] and rooftop area data[37], the area available for solar PV deployment can be estimated. This is then converted to potential solar PV capacity assuming 1.5MWp/ha for ground-mounted and floating PV, 23% panel efficiency, 40%/15% utilization rate for rooftop/façade PV, and 58Wp/m$^2$ for agrivoltaics (average value based on existing projects in Japan[38,39]). Hourly meteorological data is downloaded for selected sites for modelling. Statistical summary of the solar PV data for each service area is presented in the **Supplementary Information A.1**.

Due to limited shallow water around Japan[40], floating offshore wind turbines will be largely deployed. Recent developments of wind turbines with floating foundations make it possible to access far larger wind resources in water up to 1000m deep[29]. Globally, 66 Megawatts (MW) of floating offshore wind had been installed by 2019, of which 19 MW is in Japan[41]. Four offshore wind promotion zones were announced by the Japanese government in 2020[21], along with the first auction for floating offshore wind farms (maximum 21 MW) in the Goto sea area[42]. Floating wind capacity is expected to increase rapidly in the next decade based on announced projects in the pipeline[43]. Rapidly increasing deployment will reduce costs to a more competitive level, which will in turn lead to wider deployment.

A scoring matrix (Table 2) is used to allocate an offshore wind suitability score for every 300m * 300m cell in Japan's exclusive economic zone[44]. This suitability score represents the overall suitability of installing offshore wind turbines in a certain location. In general, sites with high average wind speeds, shallow water, reasonable distance to coast and low fishing activities are preferred. Sites that are too close to the coast (<25 km) may be affected by bird activities and may have visual impacts while sites that are too far (>1000 km) would lead to high connection costs. 100km is expected to be the optimal distance to coast. Places that are close to protected areas or ferry routes and those with high fishing activities should also be avoided to minimize disruptions to the environment and commercial activities.

*Table 2: Offshore wind scoring matrix*

| Constraint | Score | Weighting |
|---|---|---|
| Wind speed (150m)[45] | 0 at 6m/s or lower, 1 at 9m/s or higher; cubic relationship | 50% |
| Sea depth[40] | 0 at 1000m or deeper, 1 at 100m or shallower; linear relationship | 30% |
| Distance to coast[34] | 0 if distance <25km or >1000km, 0 – 1 from 25km to 100km, 1 – 0 from 100km to 1000km; linear relationship | 10% |
| Fishing hour (annual)[46] | 0 for 200 hours per year or higher, 1 for 10 hours or lower; linear relationship | 10% |
| Protected area[35] | 0 if within 1km to protected area, otherwise 1 | N/a |
| Ferry route[34] | 0 if within 1 km to ferry routes, otherwise 1 | N/a |

The total score for each cell is then calculated by:

*Equation 1*

$$\begin{aligned}\text{Total score} = &(\text{wind\_speed\_score} \times 50\% + \text{sea\_depth\_score} \times 30\% \\ &+ \text{distance\_to\_coast\_score} \times 10\% \\ &+ \text{fishing\_hour\_score} \times 10\%) \times \text{protected\_area\_score} \times \text{ferry\_route\_score}\end{aligned}$$

Indicative sites with above-average scores are selected for each service area for data download and modelling. Detailed information is available in the **Supplementary Information A.2**.

The overall GIS analysis process is summarized in Figure 1.

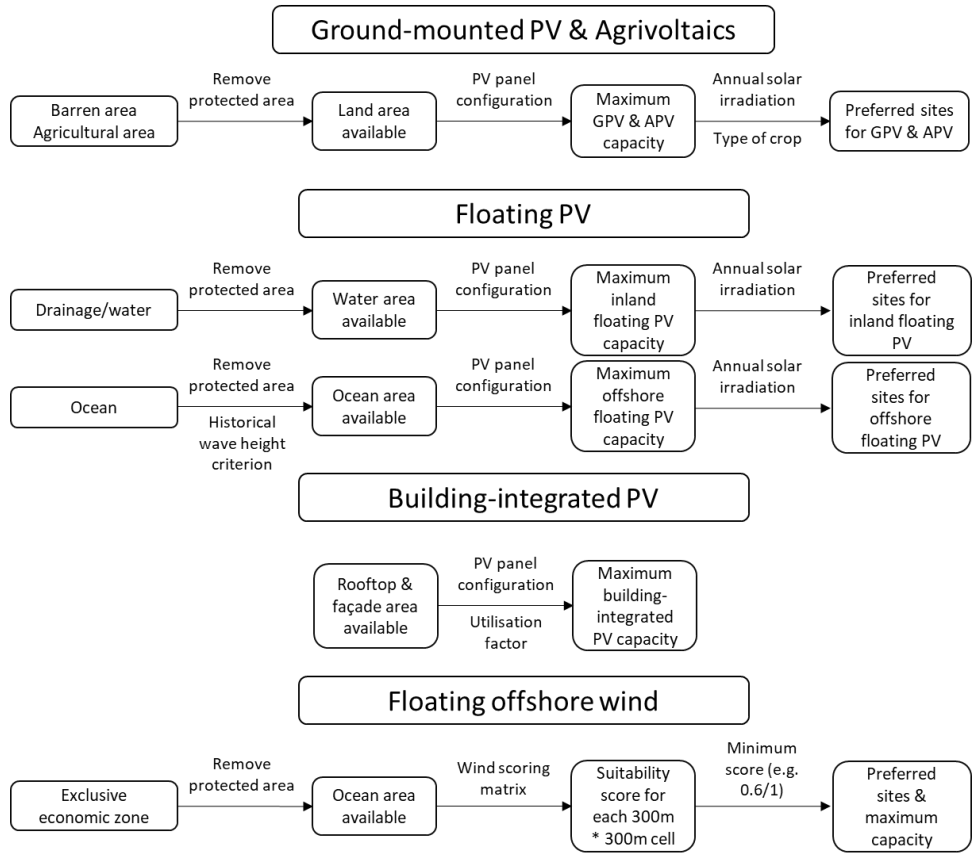

*Figure 1: GIS analysis process for the resource assessment of solar PV and offshore wind*

*2.3 Modelling scenarios*

In this study we model an interconnected Japanese electricity system in which solar PV and offshore wind supply most energy, and dispatchable generation sources (hydro, bio energy, and hydrogen) and pumped hydro energy storage provide the balance. The following primary scenarios are modelled in this study:

1. **Baseline**: electricity is supplied by solar PV, wind, hydro, biomass and hydrogen. Existing hydro and biomass capacities are used with no further expansion. A small amount of hydro (1,631 MW or 6% of total hydro capacity) is assumed to be baseload (constant output 24/7). This is determined by the minimum power from hydro over 2016-2019. The rest is assumed to be dispatchable. Bio energy is assumed to be entirely dispatchable as it is dominated by solid fuel in Japan, representing around 2% of the total electricity generation in 2019[47]. Maximum annual generation from hydro and biomass is limited to 10% of annual electricity demand, which is consistent with today's level. Other renewable energy resources (e.g., geothermal) are excluded due to the limited scale of deployment. Hydrogen (green) is assumed to be fully dispatchable and can be either imported or generated locally using curtailed electricity and are assumed to cost US$2/kg. The high voltage transmission network A in Figure 2 is used.

2. **Baseline – no hydrogen**: similar to the Baseline scenario except that dispatchable hydrogen is excluded.
3. **Wind-dominated**: similar to the Baseline scenario except that an additional energy constraint is applied, limiting generation from solar PV to 20% of annual electricity supply.
4. **Wind-dominated – no hydrogen**: similar to the Wind-dominated scenario except that dispatchable hydrogen is excluded.

In addition to these four primary scenarios, secondary scenarios are modelled to further investigate the system performance under various assumptions. All these scenarios except the 'Demand management' scenario are modelled twice – with and without dispatchable hydrogen.

5. **Nuclear**: nuclear added on top of the 'Baseline' and 'Baseline – no hydrogen' scenarios. Nuclear generation is assumed to be constant 24/7 baseload, supplying 20% of the electricity demand every year (179 Terawatt-hours (TWh) per year), which is consistent with the 2030 outlook introduced in Japan's 5th Strategic Energy Plan[48].
6. **HVDC**: similar to the 'Baseline' and 'Baseline – no hydrogen' scenarios except that high voltage transmission network B (Figure 2) is used.
7. **Okinawa-isolated**: since Okinawa is a relatively small load center and is located away from other service areas, this scenario tests whether the connection between Okinawa and other service areas is beneficial. It is similar to the 'Baseline' and 'Baseline – no hydrogen' scenarios except that the connection between Kyushu and Okinawa in high voltage transmission network A (dashed line in Figure 2) is removed.
8. **Demand management**: similar to the 'Baseline – no hydrogen' scenarios except that load shedding is allowed to take place during occasional critical periods with a fixed price of $200/MWh. Load shedding can be achieved by customers incentivized to reduce their electricity usage during peak hours, or more commonly, large industry users being paid to curtail production during a period of low solar and wind.

Demand management is not modelled on top of the 'Baseline' scenario because it has similar role in the system as hydrogen, with costs being the only difference.

9. **Current costs**: similar to the 'Baseline' and 'Baseline – no hydrogen' scenarios except that current costs of solar PV and wind (introduced in Section 2.4) are used.

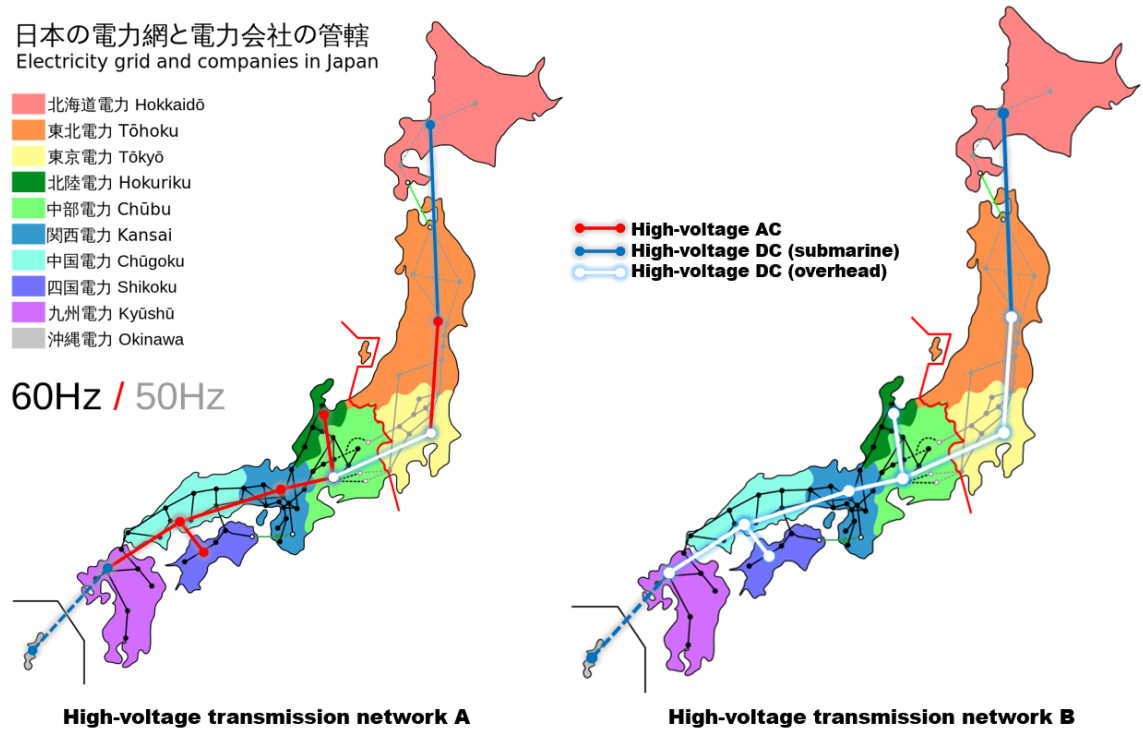

*Figure 2: Proposed high-voltage transmission network in Japan. In both networks Hokkaido-Tohoku and Kyushu-Okinawa are connected via HVDC submarine cables, and Tokyo-Chubu is connected via HVDC overhead lines due to the difference in frequencies between the two regions. In network A all other service areas are connected via HVAC lines, while in network B they are connected via HVDC overhead lines. The dashed line between Okinawa and Kyushu means in certain scenarios (Okinawa-isolated) this line is removed. Background image credit: Callum Aitchison[49].*

*2.4 Cost assumptions*

The cost assumptions used in this study are summarized in Table 2. All costs are in US dollars and assume 1 USD = 110 JPY.

*Table 3: Cost assumptions*

|  | Capital costs | Fixed O&M costs | Variable O&M costs | Purchase price | Lifetime (years) |
|---|---|---|---|---|---|
| **Solar PV current** | $2,300/kW | $49/kW p.a. | - | - | 25 |
| **Solar PV future** | $585/kW | $17/kW p.a. |  |  | 25 |
| **Floating offshore wind current** | $13,636/kW | $614/kW p.a. | - | - | 25 |
| **Floating offshore wind future** | $3,100/kW | $136/kW p.a. |  |  | 25 |
| **Pumped hydro energy storage** | $530/kW $47/kWh [a] | $8/kW p.a. $112/kW at year 20 and 40 | $0.3/MWh | - | 60 |
| **HVDC overhead** | $224/MW-km $112,000/MW [b] | $2.24/MW-km p.a. $1,120/MW p.a. [b] | - | - | 30, 50 [b] |
| **HVDC submarine** | $2,000/MW-km [c] | $20/MW-km p.a. [c] | - | - | 30 |
| **HVAC** | $1,050/MW-km [d] | $10.5/MW-km p.a. [d] | - | - | 50 |
| **Existing hydro and other renewables** | - | - | - | $100/MWh | - |
| **Existing nuclear** | - | - | - | $94/MWh | - |
| **Hydrogen via Gas Peaker** | $813/kW | $15/kW p.a. | $5/MWh | $108/MWh[e] | 20 |
| **Real discount rate** | 3.5% | | | | |

Notes:
[a] US$530/kW for power components (turbines, generators, pipes, transformers etc.), US$47/kWh for energy components (dams, reservoirs, water etc.)
[b] $/MW-km for transmission lines (50 years); $/MW for a converter station (30 years).
[c] Including transmission lines and converter stations.
[d] Including transmission lines and substations.
[e] US$108/MWh correspond to US$2/kg.

Significant cost reductions from today's level have been assumed for solar PV, floating offshore wind and hydrogen. Expected future LCOE is around US$49/MWh for solar PV, and around US$73/MWh for floating offshore wind. These cost estimations are much lower than current costs in Japan (US$104/MWh for PV[50] and US$363/MWh for floating offshore wind[51]). However, capital costs of PV and wind in Japan are high compared with comparable countries. With continued mass deployment of solar PV and wind globally, a 'global convergence' of the generation costs is likely to occur and cost reductions to global norms are expected to be achievable over the next couple of decades in multiple studies[3,51–53]. This is different from the costs of batteries, which is currently still high in most places of the world. The cost assumptions used in this study are largely in line with those in Renewable Energy Institute's 100% renewable energy study[54]. A detailed discussion of the cost assumptions adopted in this study is available in the **Supplementary Information C**.

3. Results

*3.1 Technical resource potential of PV, offshore wind and off river pumped hydro in Japan*

A total of 4,117 GW of solar PV potential has been identified using GIS analysis, comprising ground-mounted PV (3 GW), building-mounted PV (101 GW), floating PV on inland rivers, lakes and reservoirs (376 GW), floating PV on the Japanese inland sea (234 GW) and agrivoltaics (3,402 GW). The distribution of the identified solar PV potential by service area is shown in Figure 3.

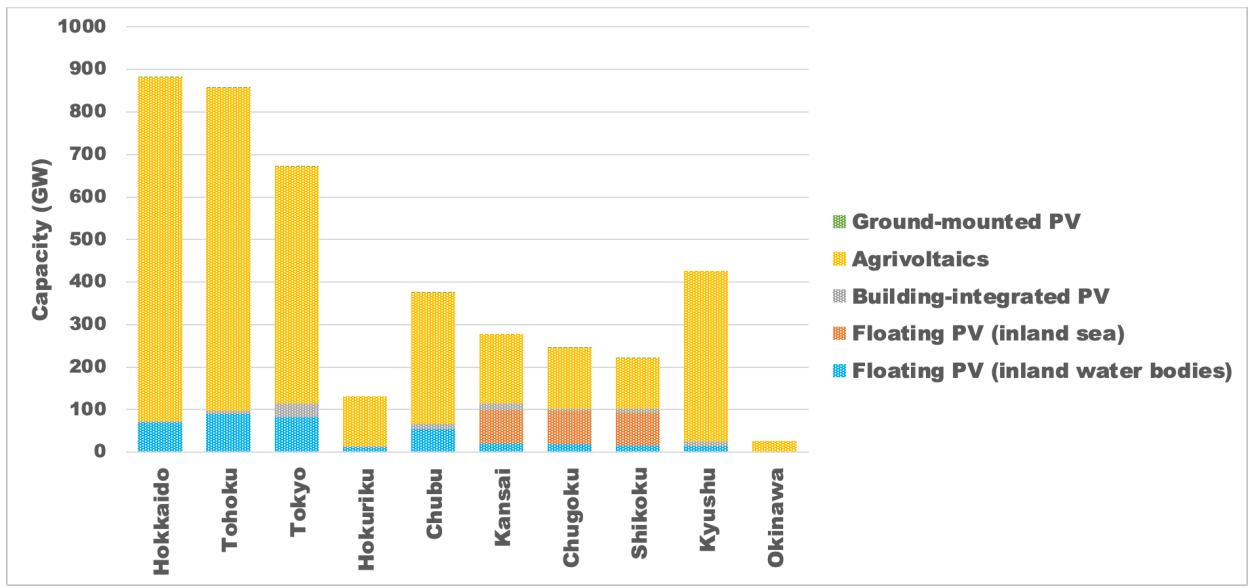

*Figure 3: Identified solar PV potential by service area. Assuming 100% utilization of the available land, water and agricultural area.*

Assuming an average capacity factor of 13%[55], the technical resource potential of solar PV in Japan represents an annual electricity generation of 4,688 TWh, which is nearly 5 times Japan's current electricity demand. Note that this assumes 100% utilization of the available land, water and agricultural area. The realistic potential of solar PV is substantially lower, because competition from other land- or water-use activities will challenge a significant fraction of the identified agrivoltaics and floating PV potential. However, the high technical resource potential identified means that only 20% of the available area need be utilized for solar PV by itself to still supply 100% electricity demand in Japan.

A vast offshore wind resource in Japan is identified. An Offshore Wind Score Map representing the relative suitability of offshore wind deployment is shown in Figure 4. Each 300m * 300m area in Japan's exclusive economic zone is scored in the range of 0 - 1 based on the local wind resource, ocean depth, distance to coast, fishing activities, protected areas and ferry routes.

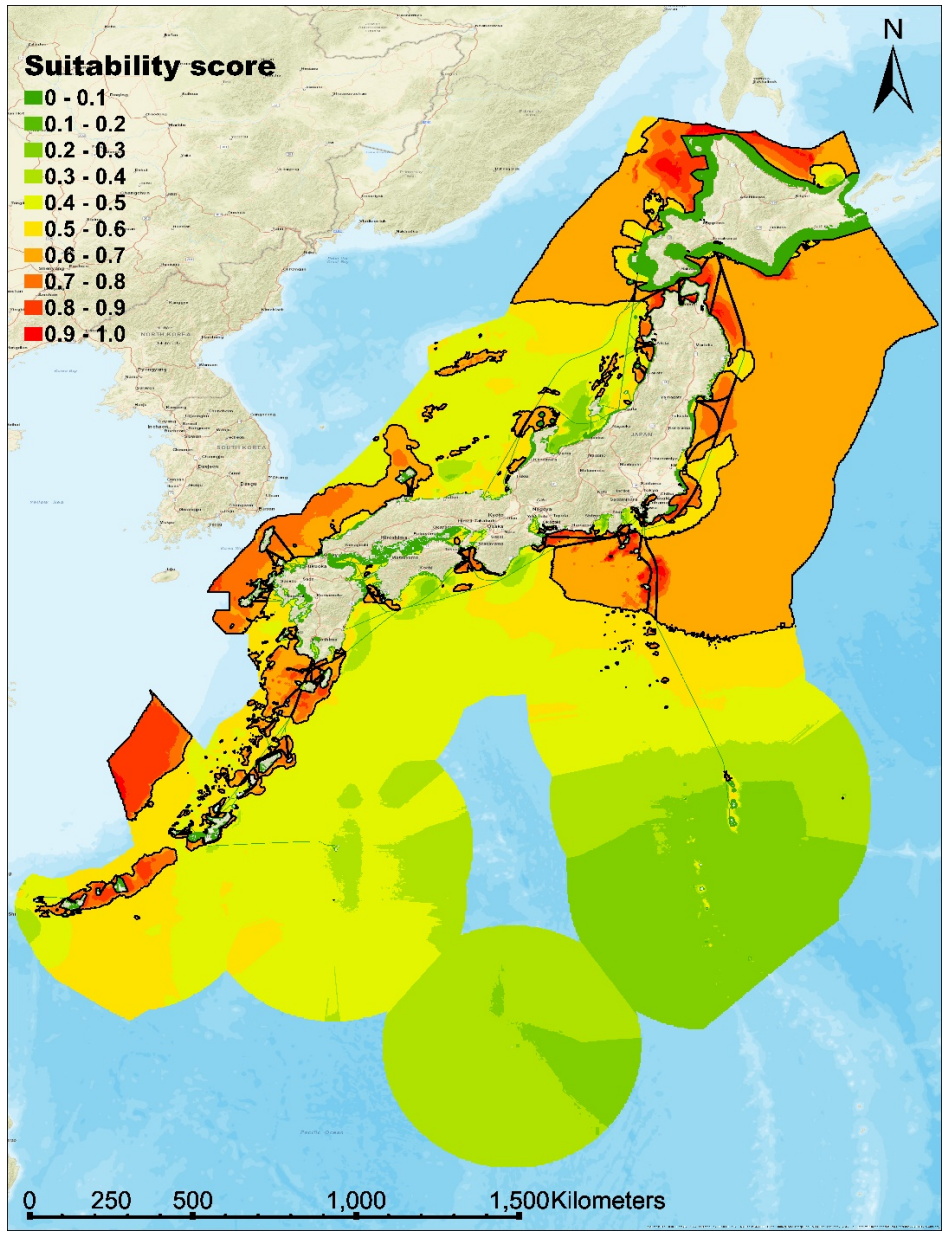

*Figure 4: Offshore wind score map. Extent: Japan's exclusive economic zone. Area enclosed by the black line represents preferable wind sites that has a suitability score above 0.6. Resolution: 300m*300m. Basemap credit: Esri.*

The estimated technical resource potential of offshore wind in Japan depends on the specified minimum score, as shown in Figure 5. An indicative wind farm with a minimum score of 0.6 is 30 km away from coast and has an annual mean wind speed of 7.5 m/s at 150 m hub height and a sea depth of 100m. This represents a typical cost-effective offshore wind site and a large area of the four offshore wind promotion zones in Japan[21] falls in this category. The sites that score higher than 0.6 represent a total offshore wind potential of 2,138 GW assuming land requirement of 5 MW/km$^2$, or 8,428 TWh annual electricity generation assuming an average capacity factor of 45%[31]. The potential capacity in each service area with a minimum score of 0.6 is shown in Figure 6.

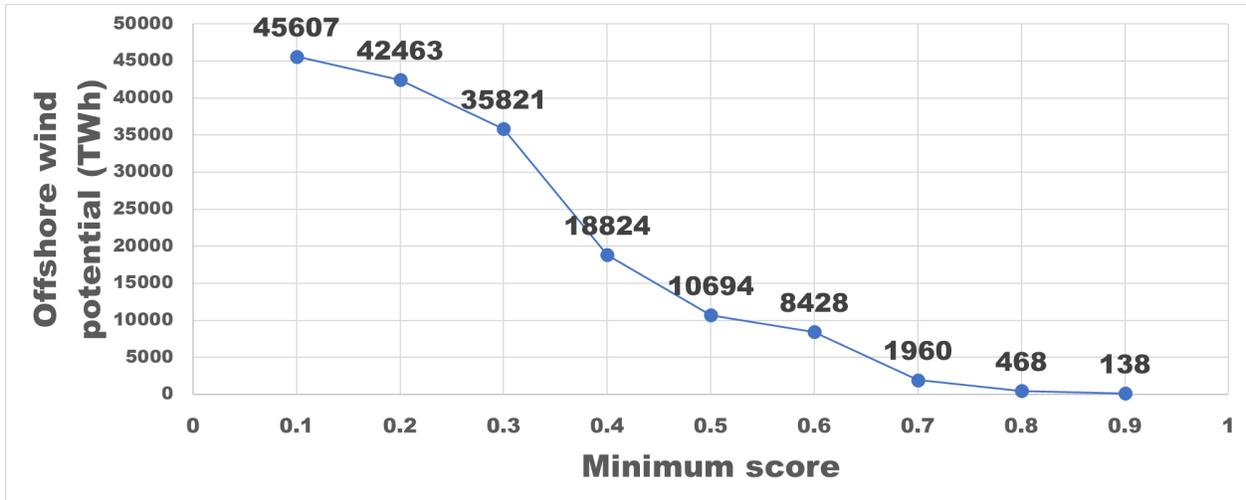

*Figure 5: Relationship between floating offshore wind potential in Japan and minimum score.*

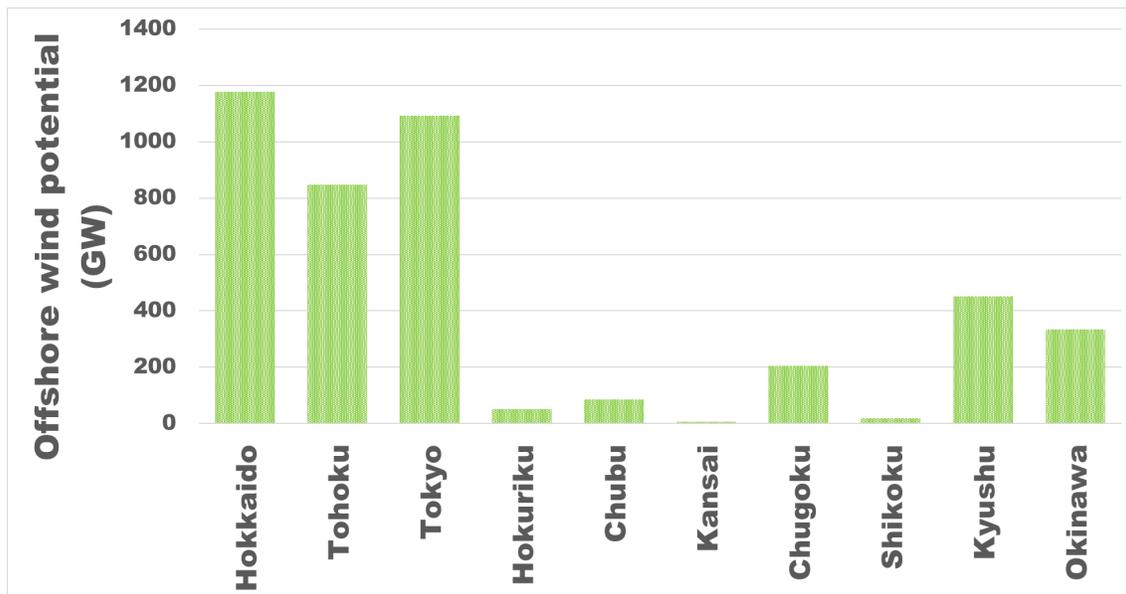

*Figure 6: Identified offshore wind potential by service area with a minimum score of 0.6.*

Many of the previous studies that attempted to investigate the offshore wind potential in Japan limited the maximum sea depth to 200m[56–58]. However, as mentioned earlier, floating offshore wind projects are already being planned at deeper water[25,26], and projects with water depth up to 1000m is technically feasible[27–29]. Excluding deep water would underestimate the offshore wind potential in Japan, as most of Japan's exclusive economic zone is in deep water. This study avoids such issues by incorporating all factors that affect the costs of the wind farms (e.g., wind speed, sea depth, distance to coast) and calculating the overall cost-effectiveness for each potential site.

A global atlas of off river pumped hydro was developed by Stocks et al.[14] It identifies 2,400 potential off river pumped hydro sites in Japan with a combined storage potential of 53 TWh. All sites are outside protected areas. The distribution of the identified sites in Japan and 3D visualization of a sample site located in Chubu is shown in Figure 7. The identified pumped hydro energy storage potential is enormous and widely distributed in most service areas except Hokkaido and Okinawa, in which only 855 GWh and 20 GWh of storage is found respectively.

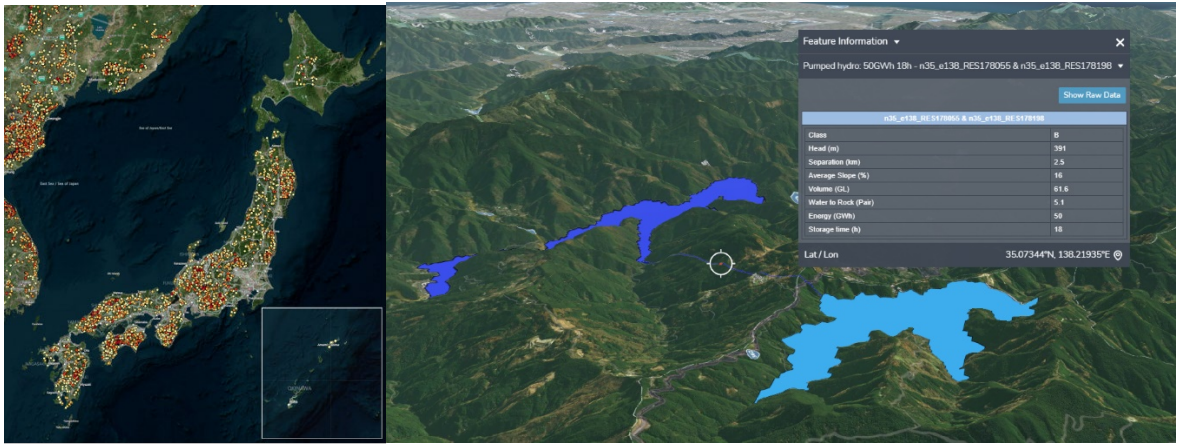

*Figure 7: Distribution of off river pumped hydro sites in Japan (left) and 3D visualization of a sample site in Chubu (right). Image from Australian Renewable Energy Mapping Infrastructure[59] and original data from the Australian National University Global Pumped Hydro Atlas[14].*

*3.2 Optimized configurations and costs of the proposed electricity system*

3.2.1 Primary scenarios – hydrogen and wind

Breakdown of LCOE, generation mix, and storage requirements for the four primary scenarios are shown in Figure 8. Detailed explanations of the scenarios can be found in Section 2.3.

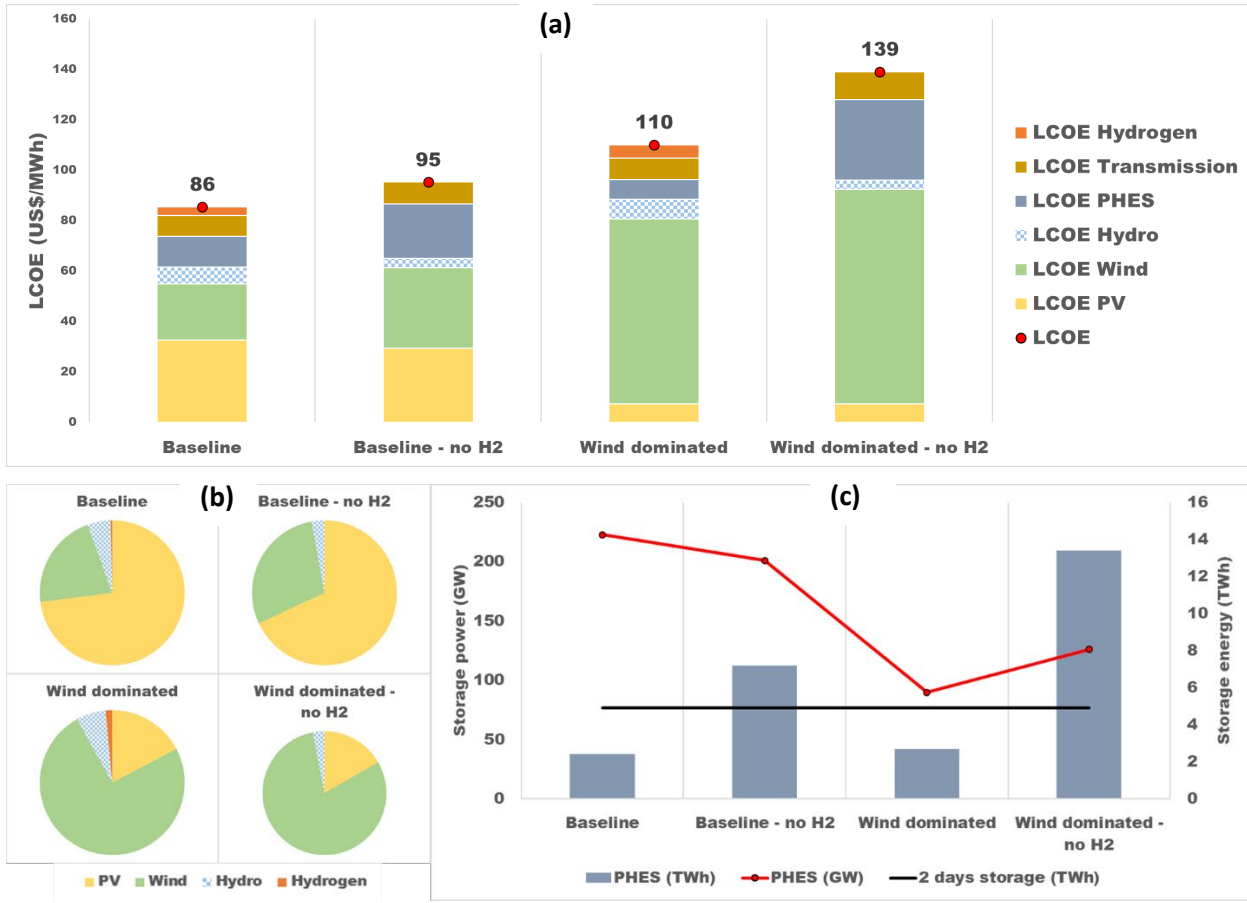

*Figure 8: Breakdown of LCOE (a), generation mix (b) and storage requirements (c) for the 4 primary scenarios. 'Hydro' refers to hydro plus biomass for simplicity. In (c) '2 days storage' represents the amount of storage that is equivalent to average electricity demand over a 2-day period (approximately 5 TWh) for comparison.*

LCOE range from US$85/MWh in the 'Baseline' scenario to US$139/MWh in the 'Wind-dominated – no hydrogen' scenario. In the two baseline scenarios solar PV is expected to supply the majority of the electricity demand, due to its lower costs. Limiting the contribution from solar PV will lead to significant increase in LCOE, as offshore wind generation is expected to cost 50% more per MWh than solar PV. It also leads to higher storage energy (TWh) but lower storage power (GW). This is because wind is more volatile than solar in Japan, and larger storage is required to accommodate occasional windless periods. However, a PV-dominated system experiences daily cycles and requires more power to store excess electricity generated during daytime.

Excluding dispatchable hydrogen has substantial impacts on the LCOE for both the 'Baseline' scenario and the 'Wind-dominated' scenario, even when the contribution from hydrogen is small from a total energy perspective (orange area in Figure 8b). The small hydrogen generation is beneficial to ride through occasional cloudy, windless periods, while most of the time it is not utilized. Most of the increase in LCOE comes from the increase in storage requirement, with the PHES component of LCOE increased by US$9/MWh for the 'Baseline' scenario and US$24/MWh for the 'Wind-dominated' scenario. This increased storage is rarely used, and it is lower cost to have gas generators on standby. Removing hydrogen leads to higher contribution from wind and lower contribution from both solar PV and existing hydro and biomass. This is because dispatchable generation is mostly utilized at night when solar is not available and storage is empty (Figure 9). The absence of hydrogen therefore requires either additional wind capacity or the storage capacity or both, with the exact mix determined by the optimization. In these scenarios both wind and storage capacities are increased for the least cost solution. The additional storage is then used in preference to hydro and biomass during periods of low generation, and therefore dispatchable hydro and biomass are needed less frequently. The lower contribution from solar PV also explains the decrease in storage power (Figure 8c) with hydrogen removed.

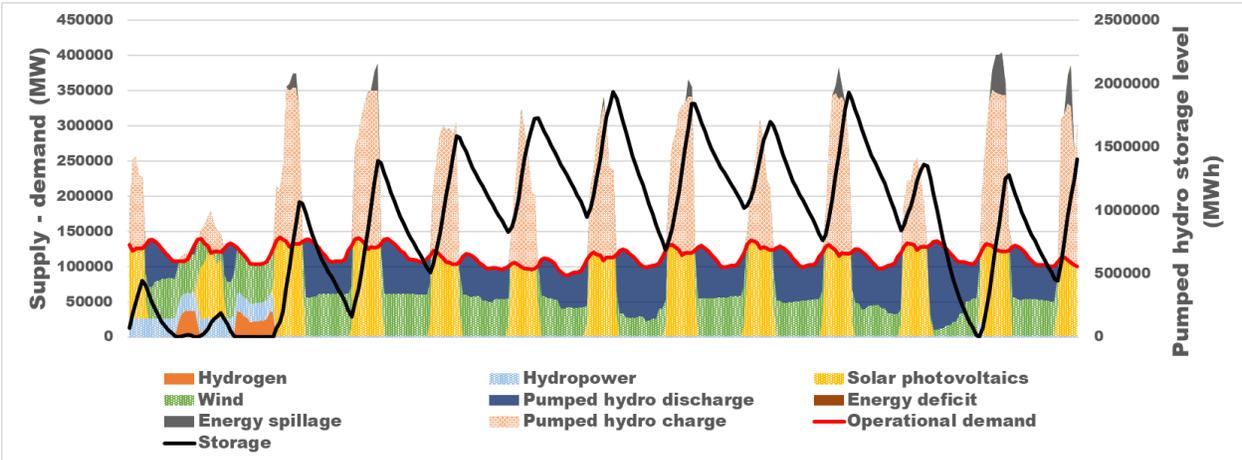

*Figure 9 : Hourly load and generation profiles over typical stressful periods for the Baseline scenario. 'Storage' refers to the state-of-charge level of the storage facilities.*

3.2.2 Secondary scenarios

Breakdown of LCOE for the secondary scenarios are shown in Figure 10 (with hydrogen) and Figure 11 (without hydrogen). The two baseline scenarios are also presented for comparison.

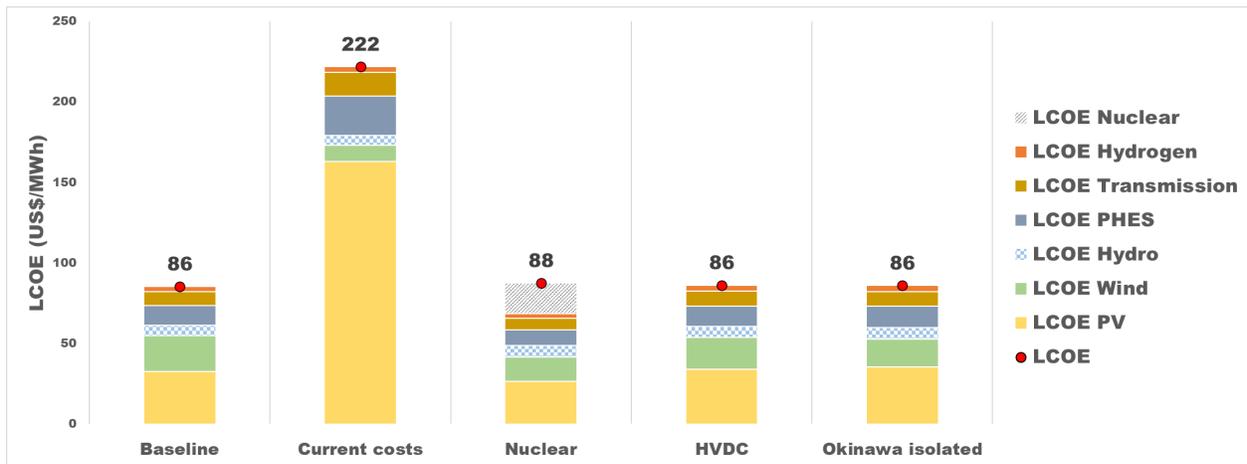

*Figure 10: Breakdown of LCOE for the secondary scenarios (with hydrogen)*

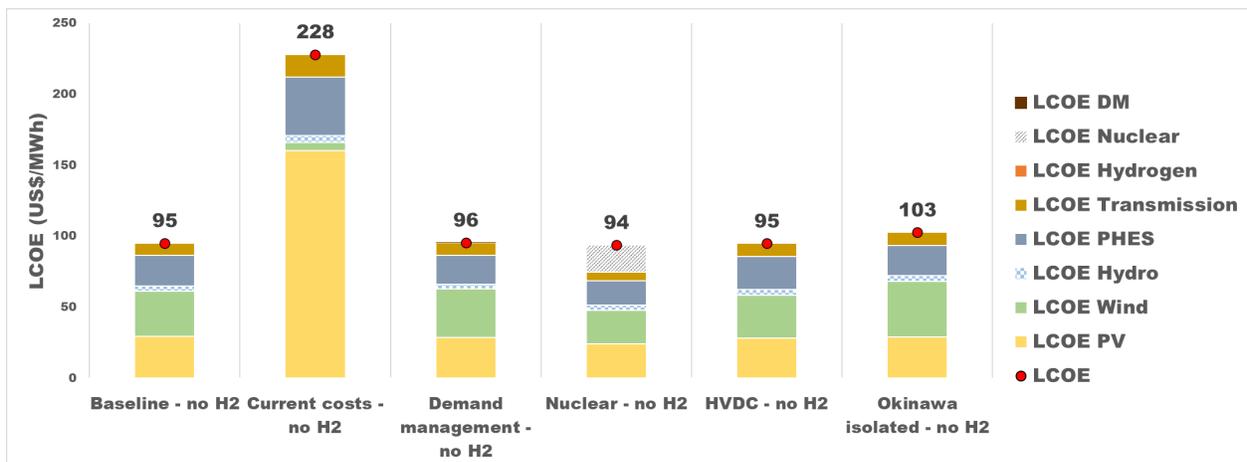

*Figure 11: Breakdown of LCOE for the secondary scenarios (without hydrogen). 'LCOE DM' represents the LCOE component due to payments for demand management.*

In both cases switching HVAC connections to HVDC yields the same results. Removing the connection between Okinawa and other regions leads to higher costs when no hydrogen is available, because the PHES resource in Okinawa is limited (only 20 GWh available) and therefore PV and wind need to be greatly overbuilt to meet the demand when generation is low (e.g., nights), which increases the LCOE in Okinawa. When flexible hydrogen is included, the LCOE is reduced to the same level when energy flow from other regions is available (i.e., the baseline scenario). Supplying 20% of the electricity demand using nuclear leads to slightly lower LCOE when no hydrogen is available, because this increase in baseload lowers the needs for PV, wind and storage. However, when hydrogen is available, increased wind plus hydrogen is lower cost than nuclear. As a result, the LCOE of the 'Nuclear' scenario is slightly higher than that of the 'Baseline' scenario.

Demand management has little impact on LCOE and is rarely used. Total load shedding over 40 years is only 3675 MWh, and during the most stressful period only 2% of the load is curtailed. This low use is due to the high price of demand management (US$200/MWh - nearly twice the cost of hydrogen). More load shedding and cost reductions may be possible if demand management can be achieved at lower prices. However, even with similar costs, demand management is unlikely to compete with hydrogen in terms of meeting demand during extended periods of low generation, because it is constrained by the number of electricity users that participate in the program and their electricity usage. For example, the average hourly load in Japan is around 100 GW, and in the 'Baseline' scenario the hydrogen capacity is 37 GW, representing nearly 40% of the load. However, curtailing 40% of the load via demand management is unrealistic.

The cost of solar PV and wind needs to approach global norms, otherwise the LCOE is more than doubled as shown in the 'Current costs' scenario. The higher cost of solar PV and wind in Japan is largely due to the lack of competition. However, prices have started to come down in recent years with more auctions for solar and wind projects and increase competition from global manufactures. Further discussions on the current and future costs of solar PV and offshore wind in Japan can be found in Section 4.2.

A detailed summary of the modelling results is shown in Table 3, including storage and transmission requirements, capacity (GW) and annual generation (TWh) for each generation technology, energy spillage and costs. Breakdown of LCOE is expressed as levelized cost of generation (LCOG) (costs of solar PV, wind, nuclear, hydro and biomass) and levelized cost of balancing (LCOB), including costs of storage, transmission, dispatchable hydrogen and spillage.

Table 4: Summary of modelling results. Detailed explanation of the scenarios can be found in the Section 2.3. '9 Area' represents the 9 service areas other than Okinawa. 'Oki' represents Okinawa. They are modelled separately in the 'Okinawa isolated' scenario. The results are calculated by the weighted average of these two models. Pumped hydro is built to have the same power capacity for pumping and generation. 'PHES (GW)' refers to this rated power capacity. 'PHES Generation (GW)' refers to the power capacity required for generation only, which is smaller than (higher power required for pumping) or equal (lower power required for pumping) to 'PHES (GW)'.

| Scenarios | Energy (TWh) | PV (GW) | PV (TWh) | Wind (GW) | Wind (TWh) | Hydro & bio (TWh) | Hydrogen (GW) | Hydrogen (TWh) | Nuclear (GW) | Nuclear (TWh) | Spillage (%) | PHES (GW) | PHES Generation (GW) | PHES (GWh) | HVDC/HVAC (GW) | LCOE (US$/MWh) | LCOG (US$/MWh) | LCOB (US$/MWh) |
|---|---|---|---|---|---|---|---|---|---|---|---|---|---|---|---|---|---|---|
| Baseline | 896 | 549 | 827 | 61 | 244 | 59 | 37 | 4 | 0 | 0 | 14% | 223 | 145 | 2415 | 530 | 86 | 61 | 24 |
| Baseline - no H2 | 896 | 496 | 763 | 87 | 326 | 32 | 0 | 0 | 0 | 0 | 14% | 201 | 152 | 7180 | 504 | 95 | 65 | 30 |
| Wind dominated | 896 | 123 | 179 | 199 | 772 | 68 | 40 | 15 | 0 | 0 | 11% | 90 | 90 | 2697 | 463 | 110 | 88 | 22 |
| Wind dominated - no H2 | 896 | 123 | 179 | 232 | 861 | 32 | 0 | 0 | 0 | 0 | 14% | 126 | 126 | 13386 | 521 | 139 | 96 | 43 |
| Current costs - no H2 | 896 | 749 | 1148 | 4 | 13 | 43 | 0 | 0 | 0 | 0 | 16% | 347 | 151 | 13750 | 924 | 228 | 171 | 57 |
| Current costs | 896 | 765 | 1146 | 6 | 22 | 52 | 38 | 2 | 0 | 0 | 18% | 406 | 152 | 5053 | 926 | 222 | 179 | 43 |
| Demand management - no H2 | 896 | 481 | 733 | 94 | 354 | 30 | 0 | 0 | 0 | 0 | 14% | 191 | 151 | 6816 | 525 | 96 | 66 | 29 |
| Nuclear - no H2 | 896 | 411 | 625 | 64 | 238 | 34 | 0 | 0 | 20 | 179 | 11% | 174 | 124 | 5574 | 390 | 94 | 71 | 23 |
| Nuclear | 896 | 446 | 680 | 42 | 168 | 62 | 31 | 3 | 20 | 179 | 12% | 181 | 125 | 2069 | 477 | 88 | 68 | 20 |
| HVDC - no H2 | 896 | 478 | 726 | 83 | 326 | 36 | 0 | 0 | 0 | 0 | 11% | 201 | 144 | 7950 | 493 | 95 | 63 | 32 |
| HVDC | 896 | 578 | 874 | 54 | 216 | 61 | 38 | 4 | 0 | 0 | 16% | 233 | 146 | 2499 | 499 | 86 | 61 | 26 |
| 9 Area no H2 | 888 | 472 | 716 | 87 | 332 | 36 | 0 | 0 | 0 | 0 | 12% | 195 | 142 | 7251 | 582 | 95 | 65 | 31 |
| 9 Area | 888 | 592 | 902 | 46 | 182 | 65 | 39 | 5 | 0 | 0 | 16% | 243 | 145 | 2469 | 557 | 86 | 60 | 26 |
| Oki no H2 | 8 | 19 | 29 | 20 | 59 | 0 | 0 | 0 | 0 | 0 | 91% | 2 | 1 | 20 | 0 | 928 | 910 | 17 |
| Oki | 8 | 6 | 9 | 0 | 0 | 0 | 1 | 1 | 0 | 0 | 12% | 2 | 1 | 20 | 0 | 79 | 41 | 38 |
| Okinawa isolated - no H2 | 896 | 491 | 746 | 107 | 391 | 36 | 0 | 0 | 0 | 0 | 18% | 196 | 143 | 7271 | 582 | 103 | 72 | 31 |
| Okinawa isolated | 896 | 598 | 911 | 46 | 182 | 65 | 40 | 6 | 0 | 0 | 16% | 246 | 146 | 2489 | 557 | 86 | 60 | 26 |

*3.3 Sensitivity analysis*

To test the impact of uncertainties in the cost assumptions, sensitivity analysis has been performed for the 'Baseline' scenario and the 'Wind-dominated' scenario. Costs of solar PV, wind, hydro, transmission, storage, hydrogen, and discount rate are varied by ±20%. To save computation time, only the 'worst year' is modelled in the sensitivity analysis. The 'worst year' contains the most stressful period in which solar irradiation and wind speeds are constantly low, which drives up the system costs. Details on how the 'worst year' is identified for different scenarios are available in the **Supplementary Information D**.

The 'worst year' is found to be 1993 for all scenarios except the two 'Wind-dominated' scenarios, for which the worst year is found to be 2006. Modelling the worst year only yields similar results compared with the full 40-year modelling for both the 'Baseline' scenario (both $85/MWh) and the 'Wind-dominated' scenario ($110/MWh vs $106/MWh), which demonstrate the effectiveness of this 'worst year' approach. The sensitivity analysis results are shown in Figure 12. For the 'Baseline' scenario, changes in LCOE are small (≤$5/MWh) when individual cost components are varied. Wind costs, PV costs, and discount rate have larger impacts on LCOE compared with other factors.

Changes in LCOE for the 'Wind-dominated' scenario are larger, with the increase in wind costs having the largest impact on LCOE ($14/MWh). It also shows an asymmetric pattern in which increased costs have larger impacts on LCOE compared with decreased costs, especially when solar PV and hydro costs are varied. This is because in the 'Wind-dominated' scenario, contribution from solar PV and hydro is constrained. Therefore, the impediment caused by moving away from the optimized configuration cannot be offset by the benefit from reducing the reliance on the 'cost-reduced' component.

The sensitivity analysis shows that the modelled LCOE depends largely on the cost assumptions adopted in this study. Given the uncertainties of future costs of generation, storage and transmission in Japan, a few dollars' difference in LCOE is within the margin of error and may have no real implications. It is therefore important to focus on the factors that have more dramatic impacts, i.e., the primary scenarios.

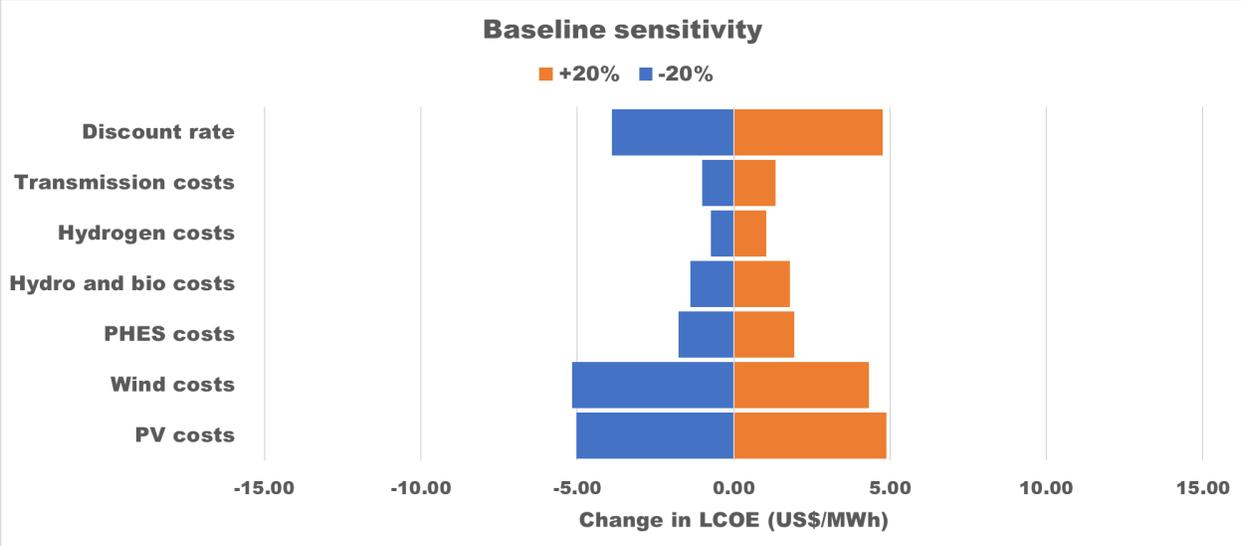

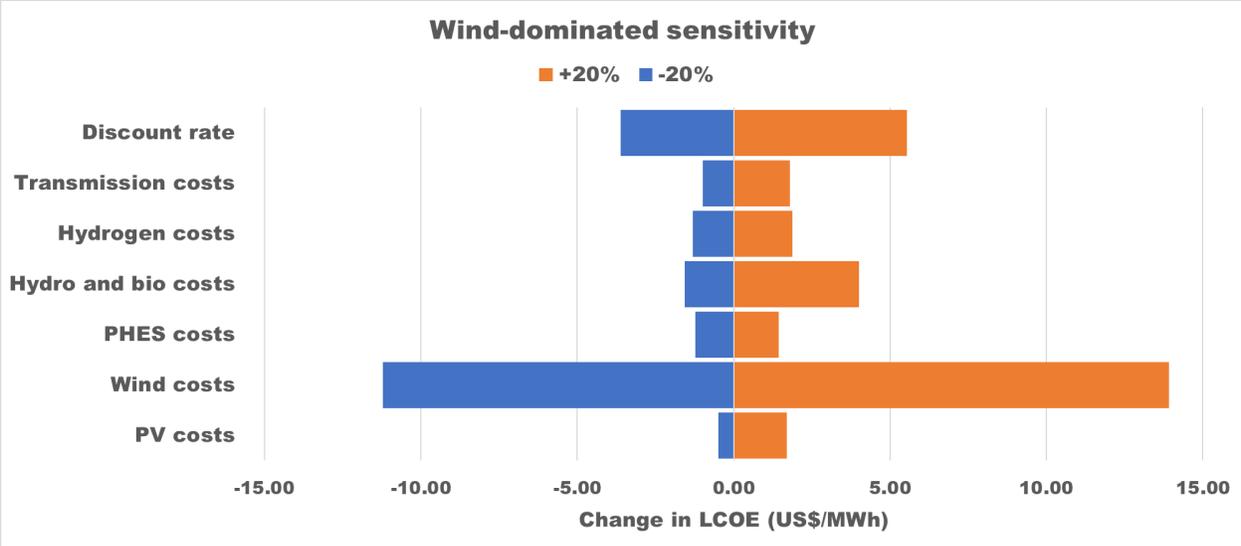

*Figure 12: Sensitivity analysis results for the 'Baseline' scenario (up) and the 'Wind-dominated' scenario (bottom). All cost components are varied ±20%.*

4. Discussion

*4.1 Japan's renewable energy resources*

In this study we have identified over 4 Terawatts (TW) of solar PV potential and over 2 TW of offshore wind potential in Japan. Combined, they represent annual generation of over 13,000 TWh, which is 14 times larger than current generation. The modelling results suggest that required PV capacity ranges from 122 GW to 765 GW, which is 3% - 19% of the identified potential. Required offshore wind capacity range from 4 GW to 232 GW, corresponding to 0.2% - 11% of the identified potential. In the 'Baseline' scenario, 549 GW of solar PV and 61 GW of wind are needed, which represent 13% and 3% of the identified potential, respectively. Note that this assumes no further expansion of existing hydro and bio capacity, and essentially all current thermal and nuclear generation are replaced by solar PV and wind. In contrast to the accepted wisdom that a maximum of 50% - 60% of the electricity demand can be realistically supplied by renewables[3], Japan has sufficient land and water area to supply its entire electricity demand with domestic renewable energy resources, mostly PV and wind. It does not have to rely on nuclear or CCS or a large amount of imported clean electricity or fuel from other countries (e.g., hydrogen from Australia or an HVDC cable from China) in the transition to a carbon neutral society. The Japan Wind Power Association is proposing 30-45 GW of domestic offshore wind generation by 2040[60]. This is more than half of the modelled wind capacity (61 GW) in the 'Baseline' scenario, although a higher capacity (87 GW) is needed in the 'Baseline – no hydrogen' scenario. Cost of offshore wind, including floating offshore wind, is expected to approach global norms in the coming decades with increased deployment and competition from global manufactures.

Even though a large technical resource potential has been identified for solar PV, as discussed earlier a significant fraction of the identified agrivoltaics and floating PV potential may not be viable due to competition from other land use activities. However, unlike PV, offshore wind turbines can effectively co-exist with other offshore activities, and therefore most of the identified offshore wind capacity is expected to be economically and socially feasible. The 'wind-dominated' scenarios are therefore included in this study based on these considerations. The maximum required wind capacity is around 232 GW ('Wind-dominated – no hydrogen' scenario), which is only 11% of the identified wind potential. In this scenario only 123 GW of solar PV is needed, nearly half of which was already deployed by 2019[61].

The storage requirement ranges from 2,069 GWh to 13,750 GWh, corresponding to approximately 1-6 days of consumption. This represents 4% - 26% of the identified storage potential. Extreme (>10,000 GWh) quantities of storage are only needed when the electricity system is dominated by a single generation technology. In the 'Baseline' scenario, 2,415 GWh or 19 GWh per million people of storage is needed to support 100% renewable electricity. This is consistent with the value (20 GWh per million people) for Australia reported in a previous study[12].

Large deployment of domestic renewable energy, especially offshore wind, in Japan may affect local lifestyle and fishery rights and as a result, may introduce social impacts. A detailed discussion on the potential impacts on local communities of the proposed pathway is out of the scope of this paper. However, in the transition to carbon neutrality, Japan might have to compromise and consider the tradeoff between the potential social impacts caused by domestic renewable energy deployment and the costs and energy security issues from importing expensive low carbon fuel or electricity.

*4.2 Levelized cost of 100% renewable electricity in Japan*

LCOE in the 'Baseline' scenario is found to be US$86/MWh, which is significantly lower than the average system prices on the spot market in Japan in 2020 (US$102/MWh[62]). Transmitting solar and wind generated electricity from Central Asia (Western China or Mongolia) to Japan via HVDC would cost about US$60/MWh assuming steady 24/7 supply, including costs of generation (US$35/MWh), costs of storage and spillage (US$18/MWh)[12] and costs of a 3000-km HVDC transmission line ($US7/MWh). The lower cost for electricity generation and balancing is due to the lower cost for onshore wind, better solar resources, and better averaging from wide distribution of resources. This is a cheaper option but there are political barriers to overcome as this may introduce energy security issues, which contradicts Japan's '3E+S' (energy security, economic efficiency, and environment plus safety) philosophy[63]. Japan can supply its electricity demand with affordable and reliable renewable energy from domestic resources, perhaps with supplementation from abroad.

If the contribution from solar PV is limited to 20% of total generation to represent the potential shortage of land for PV deployment, LCOE will increase from US$86/MWh to US$110/MWh, due to the higher cost of offshore wind. However, this is only 8% higher than the current system prices (US$102/MWh), but would effectively eliminate concerns regarding the lack of available land for PV deployment in Japan.

Dispatchable hydrogen has a large impact on the LCOE, with LCOE increased by US$9/MWh and US$29/MWh in the 'Baseline – no hydrogen' and 'Wind-dominated – no hydrogen' scenarios, respectively. Only a small amount of hydrogen is required to keep the LCOE low (0.3% - 1.5% of the total generation). This study assumes that hydrogen is imported or manufactured locally at US$2/kg (US$108/MWh) and then combusted via gas peakers. The hydrogen price assumed in this study is consistent with the projected hydrogen cost (20 yen per Nm3) in Japan's Green Growth Strategy[3]. It could be generated using curtailed electricity and then stored for use in extended periods of low generation, as the spillage of electricity (110 - 220 TWh per year) is much larger than the amount of hydrogen (2 – 15 TWh per year) needed.

A wide range of scenarios (nuclear, HVDC, Okinawa connection, demand management) result in similar costs. The highest LCOE observed comes from the 'Current costs' scenario. Extreme floating offshore wind costs (US$363/MWh, compared with cost in Europe US$132/MWh[64]) results in almost no wind deployed. Relying on PV only means that all electricity would come from pumped hydro during night, which leads to much higher storage costs. LCOG is also much higher than other scenarios, due to the high current costs of solar PV in Japan (US$104/MWh). Significant cost reduction of solar PV and offshore wind from today's level is needed to enable the cost competitiveness of the proposed renewable system.

Although still higher than global prices, a sharp decline in the costs of solar PV in Japan over recent years can be observed. Average capital cost has decreased from 372k yen/kW (US$3,382/kW) in 2013 to 253k yen/kW (US$2,300/kW) in 2020, mostly due to the decrease in module costs[50]. Continued cost reduction for solar PV in Japan is projected in a report from Renewable Energy Institute in Japan[52], in which LCOE is expected to be 5.4-5.7 yen/kWh (4.9-5.1 US cents/kWh) by 2030. On the other hand, studies show that LCOE of floating offshore wind will be cut in half between 2019 and 2032, and by 2030 the LCOE would be similar to that for fixed bottom systems at below US$100/MWh[65]. This is consistent with the findings from a study done by Bosch et al.[53], in which the average LCOE for the identified offshore wind resources in Japan is $86/MWh, with the lowest LCOE observed at deep water where floating foundations are used. Cost reductions for solar PV and offshore wind is likely to happen naturally in Japan with more solar PV and offshore wind deployed due to learning curves and increased competition. The authors are positive about significant cost reductions of solar PV and offshore wind in Japan towards global norms over the next couple of decades.

*4.3 Decarbonized energy sector*

This study focuses on the electricity sector in Japan, while to achieve carbon neutrality, other sectors also need to be decarbonized. Emissions from transport, heat and industry can be effectively eliminated by electrification, which is likely to increase the electricity demand by 30% - 50%[3]. The identified solar PV and wind potential together can supply more than 13,000 TWh of electricity per year, which, based on the estimated electricity demand presented in Table 1, is an order of magnitude more than required to supply a fully electrified energy system in Japan. The off-river pumped hydro potential in Japan is also sufficient to support higher renewable capacities, even in the most extreme scenarios. The overall effect of dispatchable hydro and biomass in meeting energy deficits during critical periods is diminished due to the limitation on further expansion. However, this can be compensated by additional dispatchable hydrogen, and the optimal contribution is likely to be higher.

Hydrogen is unlikely to compete with solar PV and wind to supply a significant fraction of electricity demand due to its higher costs (US$108/MWh excluding costs for gas peaker). However, it may be competitive as the fuel for shipping, aviation and industrial processes, as direct electrification of these processes is not available at commercial scale yet. Further work will investigate the most cost-effective decarbonization solution for each energy sector and propose an optimal pathway for the transition to a carbon neutral society.

5. Conclusion

Following Japan's commitment to carbon neutrality by 2050, this study investigates whether supplying 100% renewable electricity in Japan with domestic renewable energy resources is a feasible and affordable option. A GIS-based renewable energy resource assessment is performed, and a 40-year hourly energy balance analysis is presented. The proposed electricity system is largely supplied by solar PV and offshore wind, with intermittent generation balanced by existing hydro and biomass, dispatchable hydrogen, pumped hydro energy storage and transmission. It is found that Japan has sufficient solar PV, wind, and pumped hydro potential to support 100% renewable electricity or even 100% renewable energy. Importantly, a wide range of scenarios yield costs in the range US$84-110/MWh which are competitive with current spot prices. This gives confidence that 100% renewable electricity via solar PV, wind, and pumped hydro in Japan is workable despite uncertainties in constraints and costs. This offers an important alternative pathway to decarbonization in Japan in addition to those presented in the METI meeting[4].

57. Ushiyama, I., Nagai, H., Saito, T. & Watanabe, F. Japan's Onshore and Offshore Wind Energy Potential as Well as Long-Term Installation Goal and its Roadmap by the Year 2050. *Wind Engineering* **34**, (SAGE PublicationsSage UK: London, England, 2010).

58. Ministry of the Environment Government of Japan. Renewable Energy Potential System （リーポス（再生可能エネルギー情報提供システム））. *Ministry of the Environment Government of Japan* (2021). Available at: http://www.renewable-energy-potential.env.go.jp/RenewableEnergy/. (Accessed: 31st August 2021)

59. AREMI. ANU STORES Worldwide. *AREMI* Available at: https://nationalmap.gov.au/renewables/#share=s-9VXJcOHmnAIUMEI2. (Accessed: 17th December 2020)

60. JWPA Claims Introduction of Offshore Wind Power Over 30GW in 2040 First Meeting of Public-Private Council. *Japan Wind Power Assoication* Available at: http://log.jwpa.jp/content/0000289747.html. (Accessed: 20th December 2020)

61. *Share of renewable energy electricity in Japan, 2019 (Preliminary report)* . *Institute for Sustainable Energy Policies* (2020).

62. Trading Information: Spot Market / Intraday Market │ JEPX. *JEPX* Available at: http://www.jepx.org/english/market/index.html. (Accessed: 20th April 2021)

63. *The Electric Power Industry in Japan*. (2020).

64. Ghigo, A., Cottura, L., Caradonna, R., Bracco, G. & Mattiazzo, G. Platform optimization and cost analysis in a floating offshore wind farm. *J. Mar. Sci. Eng.* **8**, 1–26 (2020).

65. Institute, R. E. *Offshore wind in Taiwan*. (2020).
36

# Supplementary Information A: Data

Historical hourly electricity demand data is from the [Organization for Cross-Regional Coordination of Transmission Operators (OCCTO)](). However, data is available only from April 2016. In order to model 40 years of supply-demand balancing, we duplicate the demand in 2016-2019 and assume that electricity demand and load profiles in previous years are the same as those in recent years. This assumption separates weather correlation of demand but is expected to have a neglectable impact on the results, considering that the electricity demand in Japan increased from 1980 to 2015 and has been relatively flat since then. Electricity demand will likely increase in the next few decades because of renewable electrification of many energy sectors (transport, heating etc.). This will be explored in future work.

Historical hourly solar irradiation data comes from [Japan Meteorological Agency](). However, only global horizonal irradiance (GHI) is available and therefore direct normal irradiance (DNI) is calculated using the [DISC model](). Calculation results over 2007-2019 are validated with data from [Solcast](). When occasional hourly data is missing, data from the same day in the closest year is used with adjustment based on daily values. The weather data is then transferred into hourly AC output using the [System Advisor Model]() (SAM). Key SAM input assumptions are shown below:

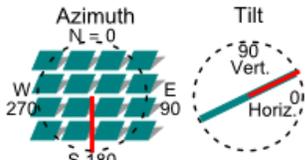

*Figure S1: SAM inputs*

Historical offshore wind generation data (170m) comes from [Windatlas.xyz](Windatlas.xyz), which is developed using [ERA5 data](ERA5 data).

## A.1 Solar data summary

*Table S1: List of PV sites*

| Service area | Pref | Site |
| --- | --- | --- |
| **Okinawa** | Okinawa | Naha |
| **Hokkaido** | Ishikari | Sapporo |
| **Tohoku** | Miyagi | Sendai |
| **Tokyo** | Tokyo | Tokyo |
| **Hokuriku** | Niigata | Niigata |
| **Chubu** | Aichi | Nagoya |
| **Kansai** | Osaka | Osaka |
| **Chugoku** | Hiroshima | Hiroshima |
| **Shikoku** | Ehime | Matsuyama |
| **Kyushu** | Fukuoka | Fukuoka |

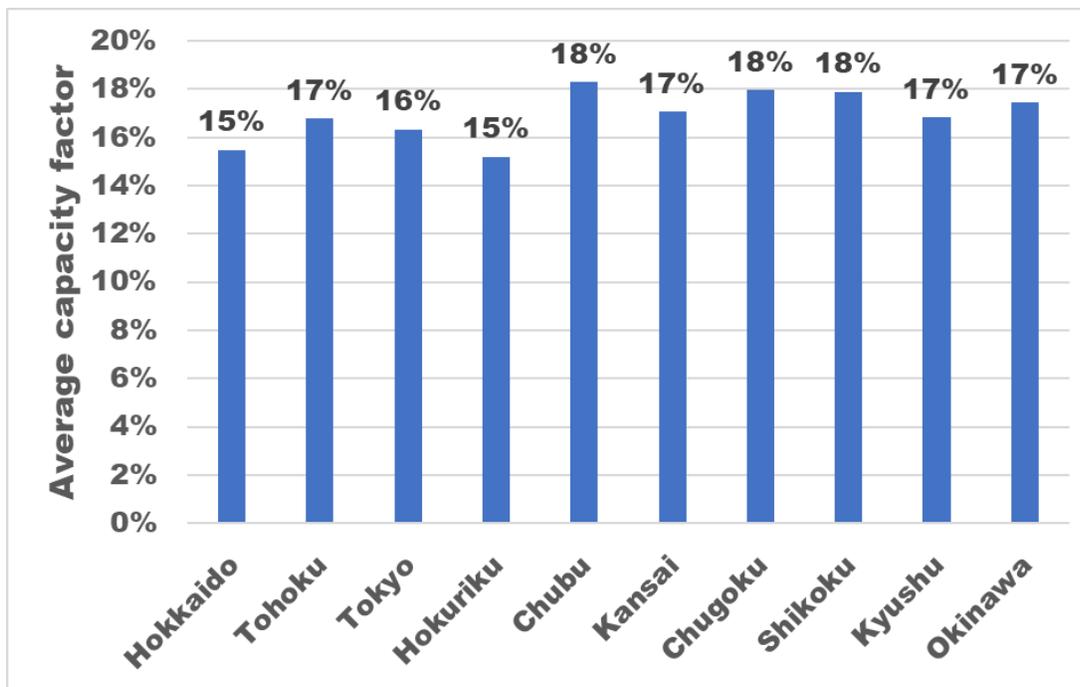

*Figure S2: PV average capacity factor by service area*

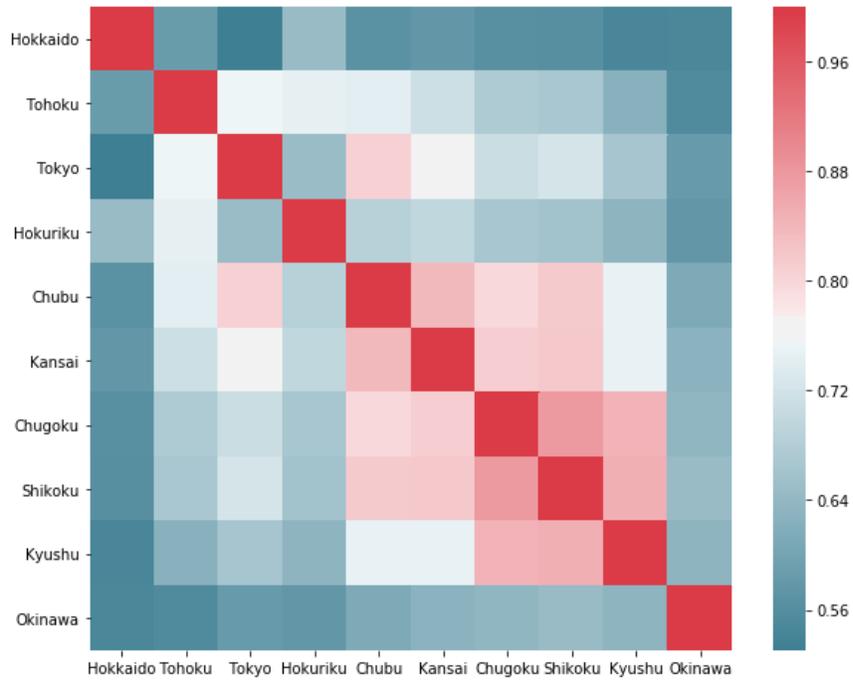

*Figure S3: PV correlation between service areas*

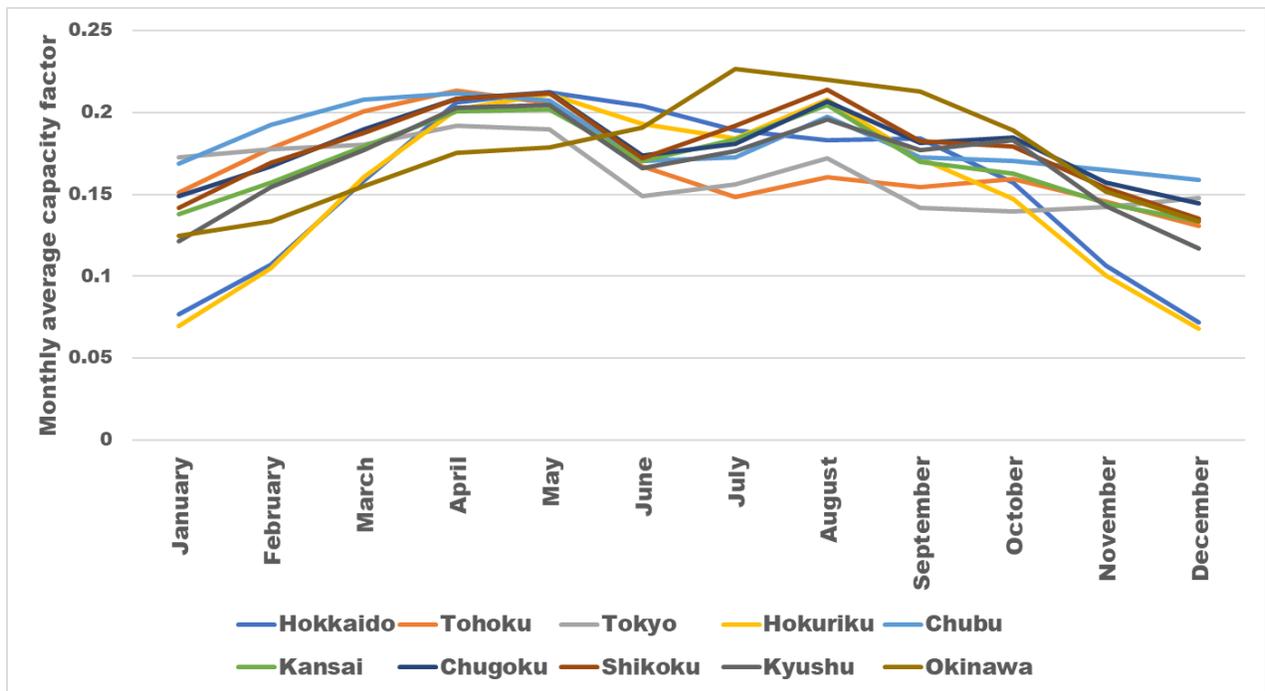

*Figure S4: PV monthly average capacity factor by service area*

## A.2 Wind data summary

*Table S2: List of offshore wind sites*

| Lat | Lon | Area | GWA_ws | Wind_class |
|---|---|---|---|---|
| 35.35 | 141.2 | Tokyo | 8.54 | 1 |
| 33.59 | 140.23 | Tokyo | 9.07 | 1 |
| 33.93 | 138.37 | Tokyo | 9.35 | 1 |
| 40.08 | 143.07 | Tohoku | 9.71 | 1 |
| 39.17 | 138.82 | Tohoku | 7.85 | 2 |
| 37.66 | 141.67 | Tohoku | 8.42 | 2 |
| 33.06 | 134.81 | Shikoku | 8.31 | 2 |
| 27.74 | 125.62 | Okinawa | 8.58 | 1 |
| 25.46 | 126.21 | Okinawa | 8.26 | 2 |
| 36.79 | 134.56 | Kansai | 7.61 | 2 |
| 33.39 | 128.33 | Kyushu | 8.06 | 2 |
| 34.85 | 130.31 | Kyushu | 7.84 | 2 |
| 32.1 | 128.82 | Kyushu | 7.86 | 2 |
| 37.61 | 136.33 | Hokuriku | 7.81 | 2 |
| 39.3 | 134.92 | Hokuriku | 8.02 | 2 |
| 41.83 | 145.24 | Hokkaido | 9.88 | 1 |
| 44.9 | 144 | Hokkaido | 8.7 | 1 |
| 44.67 | 140.31 | Hokkaido | 8.69 | 1 |
| 37 | 132.96 | Chugoku | 7.83 | 2 |
| 35.42 | 131.1 | Chugoku | 7.9 | 2 |
| 33.92 | 137.87 | Chubu | 9.23 | 1 |

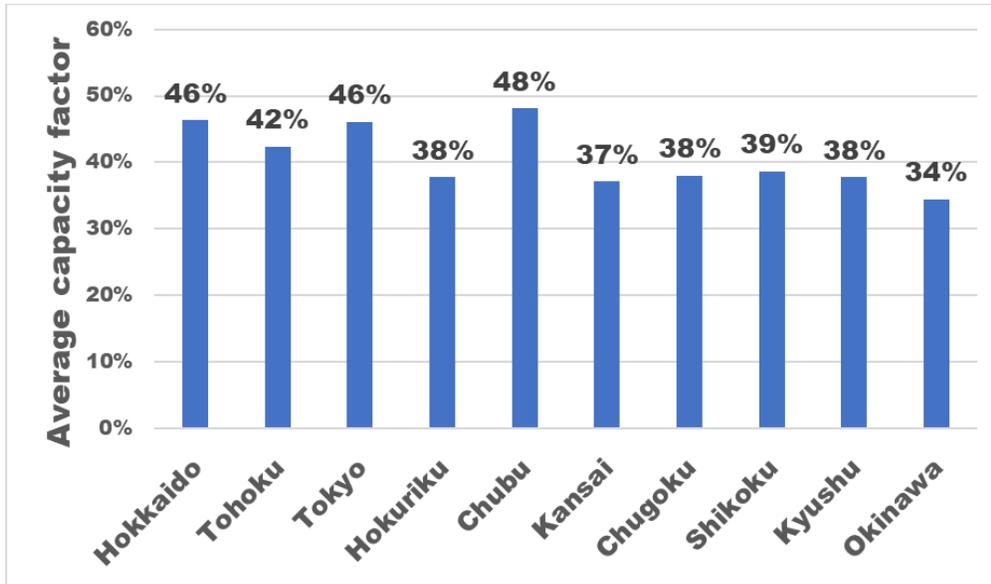

Figure S5: Average offshore wind capacity factor by service area

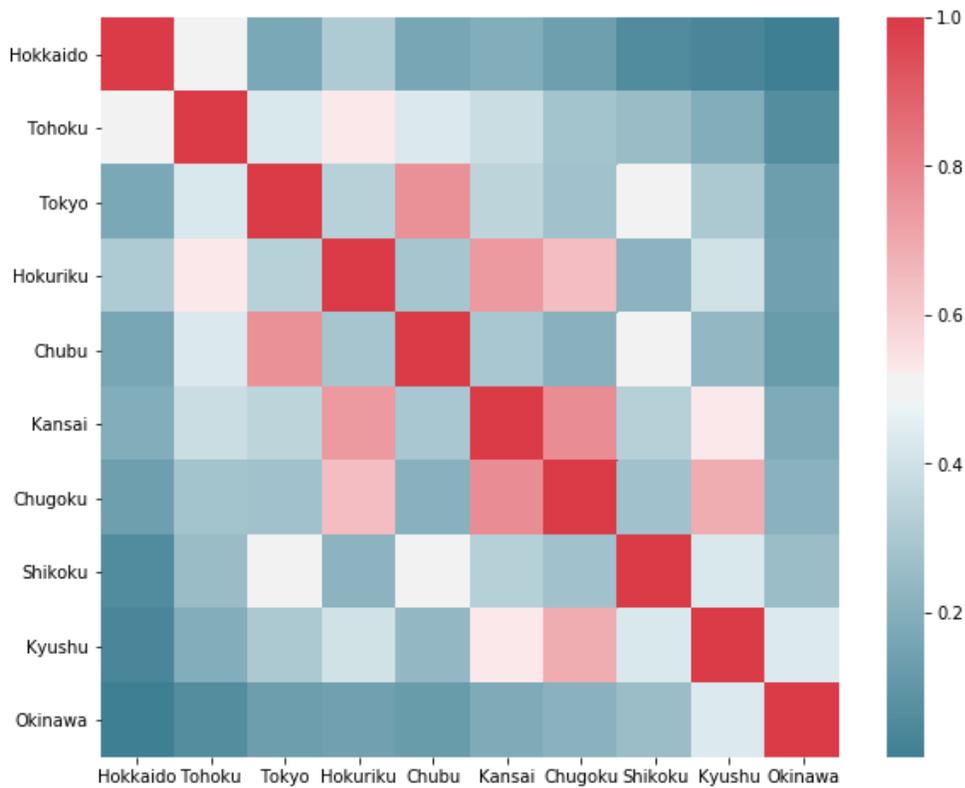

Figure S6: Offshore wind correlation between service areas

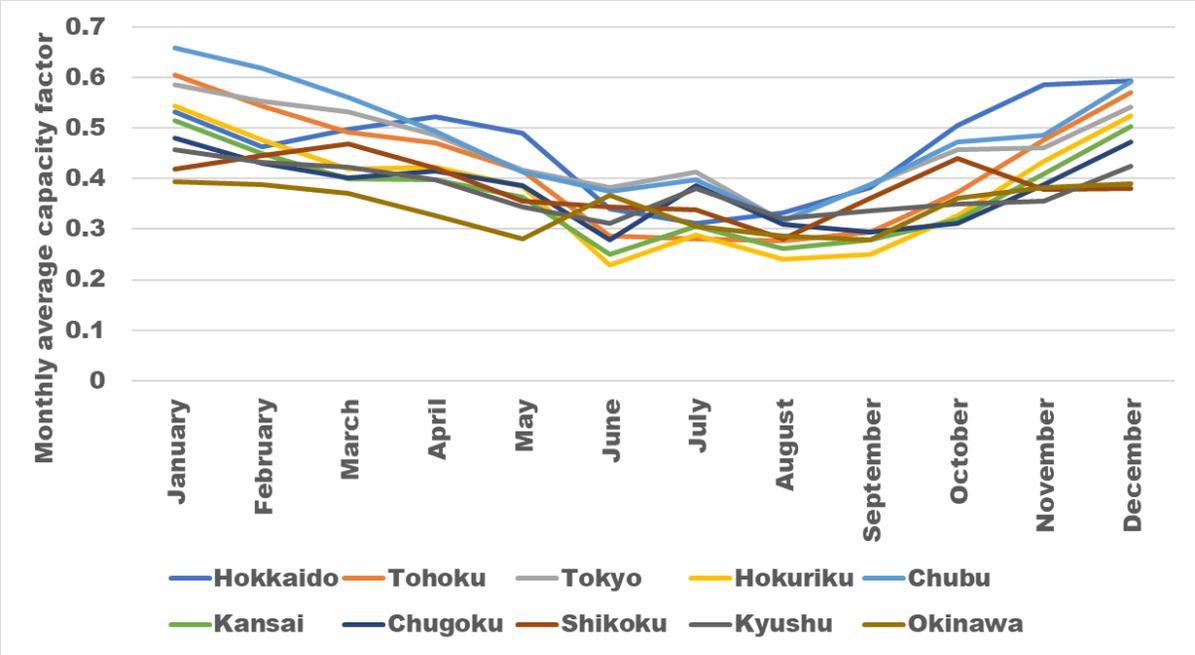

*Figure S7: Offshore wind monthly average capacity factor by service area*

# Supplementary Information B: Model formulation

This study uses a modified version of the modelling framework introduced by Lu et al. The model is implemented in Python and uses differential evolution to identify the least-cost electricity system configuration that meets defined reliability, resource, energy and transmission constraints. These constraints are defined as follows:

- **Reliability constraint**: electricity generation must meet demand in every timestep unless a specified amount of deficit is allowed, to represent load shedding in certain scenarios.
- **Resource constraint**: installed capacity of a technology in a service area must not exceed the identified technical resource potential of this technology in this service area.
- **Energy constraint**: total generation from a certain technology must not exceed the specified maximum generation from this technology.
- **Transmission constraint**: power flow in a transmission line must not exceed the specified maximum capacity of this transmission line.

This section will firstly introduce the algorithm of the model in detail, and then introduce the modifications made for this study.

For a given set of optimization parameters (e.g., PV, wind and storage capacity in each service area) with defined upper bounds determined by the resource constraint, the model uses differential evolution to find the parameters with which an objective function returns the lowest value. The objective function consists of the LCOE calculated based on energy balance simulations, and the penalties for not meeting the reliability, energy, or transmission constraints.

The model consists of the following modules:

- **Input**: import model inputs and assumptions, including time series of electricity demand and meteorological data, cost assumptions, energy constraints, resource constraints etc.

- **Simulation**: for a given set of optimization parameters (PV, wind, and storage capacity), simulate long-term hourly energy-demand balance in a perfectly interconnected system (aggregation of generation, demand, and storage in all service areas). Hourly generation from solar PV and wind are calculated based on the defined generation capacity and hourly meteorological data. Hourly generation from hydro and other renewables is an input parameter for this module (discussed later in the 'Optimization' module). Hourly net load is calculated using the following equation:

*Equation 1*

$$Netload(t) = Demand(t) - GPV(t) - GWind(t) - GHydro(t).\ t = 1,2,\dots 350399, 350400$$

Where $GPV(t)$, $GWind(t)$, $GHydro(t)$ refers to the generation from PV, wind, and hydro (and other renewables) respectively. Positive netload means generation cannot meet demand therefore energy from storage is required. Negative netload means excess electricity is available and can be stored.

Charging and discharging of storage are then calculated by:

*Equation 2*

$$Charge(t) = \min\left(-1 \times \min(0, Netload(t)), Pcapacity, \frac{Scapacity - Storage(t-1)}{efficiency}\right)$$

*Equation 3*

$$Discharge(t) = \min\left(\max(0, Netload(t)), Pcapacity, Storage(t-1)\right)$$

Where $Pcapacity$ and $Scapacity$ are the power (GW) and energy (GWh) capacities of storage. $Storage(t-1)$ is the storage level in the previous hour. $Efficiency$ is the roundtrip efficiency of storage. For simplicity discharging is assumed to be 100% efficient and all loss occurs during charging.

When the netload is positive and larger than either the energy available in storage or the storage power capacity, discharge is smaller than netload and demand cannot be met. In this case energy deficit exists:

*Equation 4*

$$Deficit(t) = \max(0, Netload(t) - Discharge(t))$$

When the netload is negative and its absolute value is larger than the storage power capacity or the difference between the maximum storage level and current storage level, charge is smaller than absolute value of netload and energy spillage exists:

*Equation 5*

$$Spillage(t) = -1 \times \min(0, Netload(t) + Charge(t))$$

Storage level is then adjusted accordingly based on charging and discharging:

*Equation 6*

$$Storage(t) = Storage(t-1) - Discharge(t) + Charge(t) \times efficiency$$

Hourly profiles are then exported to be used later in other modules.

- **Transmission**: simulate electricity transmission based on the results from the 'Simulation' module. Outputs from the 'Simulation' module are system-wide hourly profiles, which are scaled to create regional profiles based on the capacities in each region. Required energy import is then calculated for each region and for each hour:

*Equation 7*

$$Import(t,j) = Load(t,j) + Charge(t,j) + Spillage(t,j) - PV(t,j) - Wind(t,j) - Hydro(t,j) - Discharge(t,j) - Deficit(t,j)$$

Where $t$ represents the current timestep and $j$ represents the current region (service area).

Electricity transmission between regions is then determined by each region's need for energy import. For example, electricity transmission between Kyushu and Okinawa equals Okinawa's required energy import, while transmission between Tohoku and Tokyo depends not only on Tohoku's energy import but also transmission between Hokkaido and Tohoku (refer to Figure 2 in the main paper for the proposed transmission network).

For each line, its transmission capacity (power) is determined by the maximum electricity transmitted in any hour. This is then used to calculate system costs in the 'Optimization' module.

- **Optimization**: implement differential evolution to calculate the objective function for different sets of optimization parameters. The objective function consists of the Levelized cost of Electricity (LCOE) calculated based on the results from the Simulation and Transmission modules, and the penalties (heavily penalized by multiplying the results by $10^6$ as shown below) for not meeting the reliability, energy or transmission constraints:

*Equation 8*

$$F(x) = LCOE + P_{deficit} + P_{energy} + P_{transmission}$$

Where:

*Equation 9*

- $$LCOE = \frac{Cost_{PV} + Cost_{wind} + Cost_{transmission} + Cost_{storage} + Cost_{hydro}}{annual\ electricity\ demand}$$

*Equation 10*

- $P_{deficit} = max(0, 10^6 \times (\sum_{t=0}^{n}(Deficit(t)) - allowance));$
    - $n$: number of timesteps (e.g., $n = 87600$ for 10-year hourly modelling).
    - $allowance$: specified allowable deficit due to demand shedding

*Equation 11*

- $P_{energy} = max(0, 10^6 \times \sum(\sum_{t=0}^{n} G(i,t) - GMax_i))$;
    - $G(i,t)$: generation from technology $i$ (e.g., hydro) at timestep $t$;
    - $GMax\_i$: specified maximum generation from technology $i$

*Equation 12*

- $P_{transmission} = max(0, 10^6 \times \sum(CT_k - CTMax_k))$;
    - $CT_k$: calculated required capacity of transmission line k
    - $CTMax_k$: specified maximum capacity of transmission line k

As mentioned earlier one of the inputs parameters for the 'Simulation' module is the hydro generation profile. In the 'Optimization' module, 'Simulation' is performed twice with different hydro generation profiles:

- 1st Simulation: hydro = baseload 24/7. This assumes only run-off-river hydro is used for generation. Energy deficit is calculated and is assumed to be filled by flexible hydro. Total hydro generation ($GHydro$) is then calculated by the sum of run-off-river generation and flexible generation required to fill this deficit. This total generation is used to calculate $P_{energy}$ for hydro.
- 2nd Simulation: hydro = maximum capacity 24/7. This assumes all hydro runs at full capacity continuously. Energy deficit is calculated again and is expected to be zero. $P_{deficit}$ is calculated.

After these two 'Simulations', the 'Transmission' module runs, and required transmission capacities are estimated. LCOE is then calculated for this set of optimization parameters. When the 'Optimization' module finishes an optimization system configuration that results in the lowest LCOE as well as zero penalty is generated.

- **Dispatch**: determine the actual generation profile for flexible hydro. In the first simulation all deficit is expected to be filled by flexible hydro. However, the hourly deficit profile cannot be directly used as flexible hydro generation profile because deficit may exceed hydro capacity. The 'Dispatch' module starts with a hydro profile that operates at maximum capacity continuously, which is expected to result in zero deficit. Then for each hour of the modelling period, flexible hydro is turned off so that the hydro generation is from run-off-river hydro only. The 'Simulation' model is used to calculate deficit again. If the deficit is still zero then flexible hydro is turned off for this hour, otherwise it is turned on again. By the end of the 'Dispatch' module, a hydro generation profile is created while making sure the reliability constraint is still met.
- **Statistics**: check the feasibility of the solution and export energy balance profiles and statistics.

In this study, the following modifications are made to better represent the electricity system in Japan:

- Incorporate flexible hydrogen as another method for balancing variable PV and wind generation. In addition to PV, wind, and storage capacities, hydrogen capacity in each service area is added to the optimization parameters. Hydrogen is also added to the 'Simulation' and 'Transmission' modules and works in the same way as flexible hydro. However, additional simulations are performed to estimate the annual hydrogen generation in the 'Optimization' module.
    - 1$^{st}$ Simulation: hydro set to baseload (run-off-river) and hydrogen set to zero for the entire modelling period. The deficit is expected to be filled by both flexible hydro and hydrogen.
    - 2$^{nd}$ Simulation: hydro set to maximum capacity and hydrogen set to zero. The deficit is expected to be filled by flexible hydrogen. $GHydrogen$ is then calculated and subtracted from the deficit in the 1$^{st}$ simulation to calculate $GHydro$. $P_{energy}$ is then calculated for both hydro and hydrogen using Equation 11 for energy constraint.

- o 3$^{rd}$ Simulation: both hydro and hydrogen are set to maximum capacity. Deficit is calculated and is expected to be zero. $P_{deficit}$ is calculated for reliability constraint.
- Incorporate nuclear as an additional baseload in the 'Nuclear' scenarios. Nuclear is expected to supply 20% of the electricity demand in the 'Nuclear' scenarios. This baseload is incorporated into the model in the same way as run-off-river hydro and is added on top of all Simulations. For example, input parameters for the 1$^{st}$ Simulation are run-off-river hydro, zero hydrogen, and nuclear baseload.
- Refine the 'Dispatch' module. The original 'Dispatch' module is computationally intensive, as the 'Simulation' model needs to run for every hour of the modelling period, corresponding to 40*8760 = 350,400 runs for this study. Therefore, in the study the 'Dispatch' module is largely refined to a deficit targeted model. The model iterates through each value in the hourly deficit profile calculated from the 2$^{nd}$ simulation and attempts to calculate the expected hydrogen generation in that hour. If the deficit cannot be fully met by the real-time hydrogen, either because the maximum hydrogen capacity is reached in that hour or the annual hydrogen generation reaches the limit, the model will increase the hydrogen generation in the hours ahead to the required level (subject to capacity and annual generation limit) until the remaining deficit is met. In another word, hydrogen is used to pre-charge storage. By the end of this process, all deficit from the 2$^{nd}$ Simulation is expected to be met by flexible hydrogen and a hydrogen profile is generated. It is then modelled together with run-off-river hydro to calculate the deficit that needs to be filled by flexible hydro. The same process is repeated until all deficit is filled and a hydro generation profile is generated. The entire process is illustrated in the figure below.

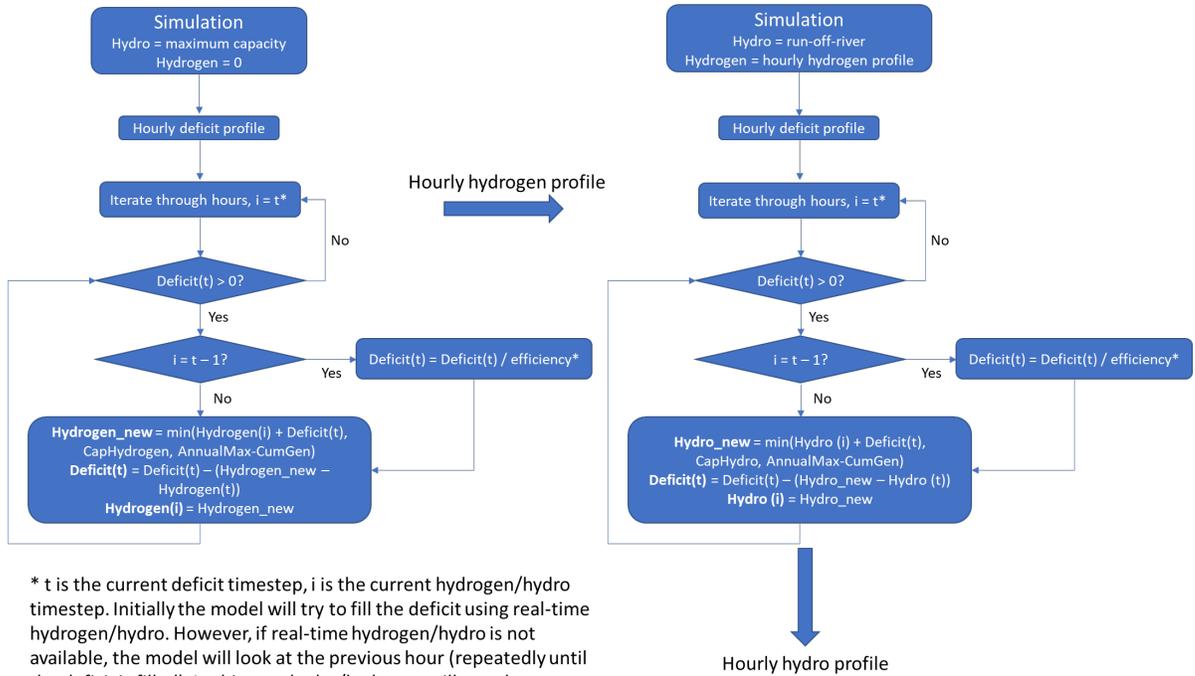

*Figure S8: Flow chart for the modified 'Dispatch' module*

# Supplementary Information C: Cost assumptions

Assumes 1 USD = 110 JPY.

*Table S3: Cost assumptions*

|  | Capital costs | Fixed O&M costs | Variable O&M costs | Purchase price | Lifetime (years) |
|---|---|---|---|---|---|
| **Solar PV current** | $2,300/kW | $49/kW p.a. | - | - | 25 |
| **Solar PV future** | $585/kW | $17/kW p.a. |  |  | 25 |
| **Floating offshore wind current** | $13,636/kW | $614/kW p.a. | - | - | 25 |
| **Floating offshore wind future** | $3,100/kW | $136/kW p.a. |  |  | 25 |
| **Pumped hydro energy storage** | $530/kW $47/kWh [a] | $8/kW p.a. $112/kW at year 20 and 40 | US $0.3/MWh | - | 60 |
| **HVDC overhead** | $224/MW-km $112,000/MW [b] | $2.24/MW-km p.a. $1,120/MW p.a. [b] | - | - | 30, 50 [b] |
| **HVDC submarine** | $2,000/MW-km [c] | $20/MW-km p.a. [c] | - | - | 30 |
| **HVAC** | $1,050/MW-km [d] | $10.5/MW-km p.a. [d] | - | - | 50 |
| **Existing hydro and other renewables** | - | - | - | US$100/MWh | - |
| **Existing nuclear** | - | - | - | US$94/MWh | - |
| **Hydrogen via Gas Peaker** | $813/kW | $15/kW p.a. | $5/MWh | US$108/MWh [e] | 20 |
| **Real discount rate** | 3.5% | | | | |

*Notes:*

[a] US$530/kW for power components (turbines, generators, pipes, transformers etc.), US$47/kWh for energy components (dams, reservoirs, water etc.)

[b] $/MW-km for transmission lines (50 years); $/MW for a converter station (30 years).

[c] Including transmission lines and converter stations.

[d] Including transmission lines and substations.

[e] US$108/MWh correspond to US$2/kg.

## Solar PV

According to a report published by Renewable Energy Institute (REI), average LCOE for solar PV in Japan is estimated to be 15.3 yen/kWh (13.9 US cents/kWh) as of 2018/2019. Lower prices are reported in a recent report published by Ministry of Economy, Trade and Industry (METI). In this report the 2020 prices for solar PV in Japan are 21 yen/kWh (19.1 US cents/kWh) for residential use (<10kW), 12 yen/kWh (10.9 US cents/kWh) for business use (50-250kW), and 11.44 yen/kWh (10.4 US cents/kWh) for commercial use (>250kW). Although still higher than global prices, a sharp decline in the costs of solar PV in Japan over recent years can be observed (red line in Figure S9). Average capital cost has decreased from 372k yen/kW (US$3,382/kW) in 2013 to 253k yen/kW (US$2,300/kW) in 2020, mostly due to the decrease in module costs (orange region in Figure S10). In the meantime, O&M costs are relatively stable, decreasing slightly from 5.5k yen/kW/year (US$50/kW/year) in 2019 to 5.4k yen/kW/year (US$49/kW/year) in 2020.

Continued cost reduction for solar PV in Japan is projected in the REI report, in which a bottom-up approach is used to estimate the solar PV costs in Japan in 2030. In the 'Standard' scenario in which 20% of the modules are supplied by Japanese companies with the rest supplied by international companies, LCOE is expected to be 5.7 yen/kWh (5.1 US cents/kWh) by 2030. On the other hand, in the 'Global Convergence' scenario in which *'price of Japanese manufacturers converges with that of overseas manufacturers'*, LCOE is expected to be 5.4 yen/kWh (4.9 US cents/kWh) by 2030. Both figures assume 25 years lifetime. *'Module conversion efficiency, reduction in wafer thickness, improvements in productivity, and lower raw material and manufacturing machinery costs due to market expansion'* are the main drivers for such cost reduction. In this study, cost estimates for the 2030 'global convergence' scenario are used to represent future costs of solar PV in Japan.

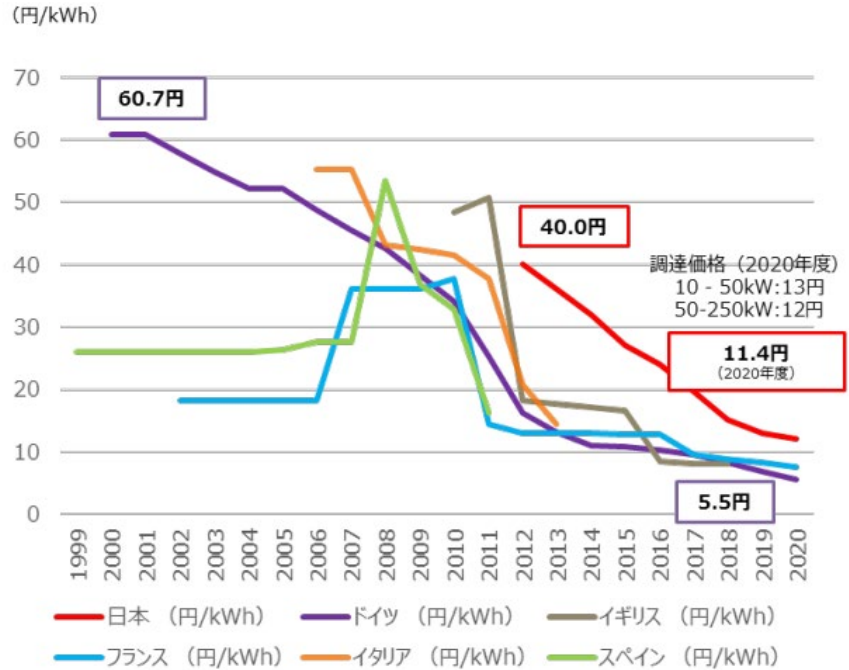

*Figure S9: Costs of solar PV: Japan and other countries. The 6 legends represent Japan, Germany, UK, France, Italy and Spain from left to right, top to bottom. Source: METI report*

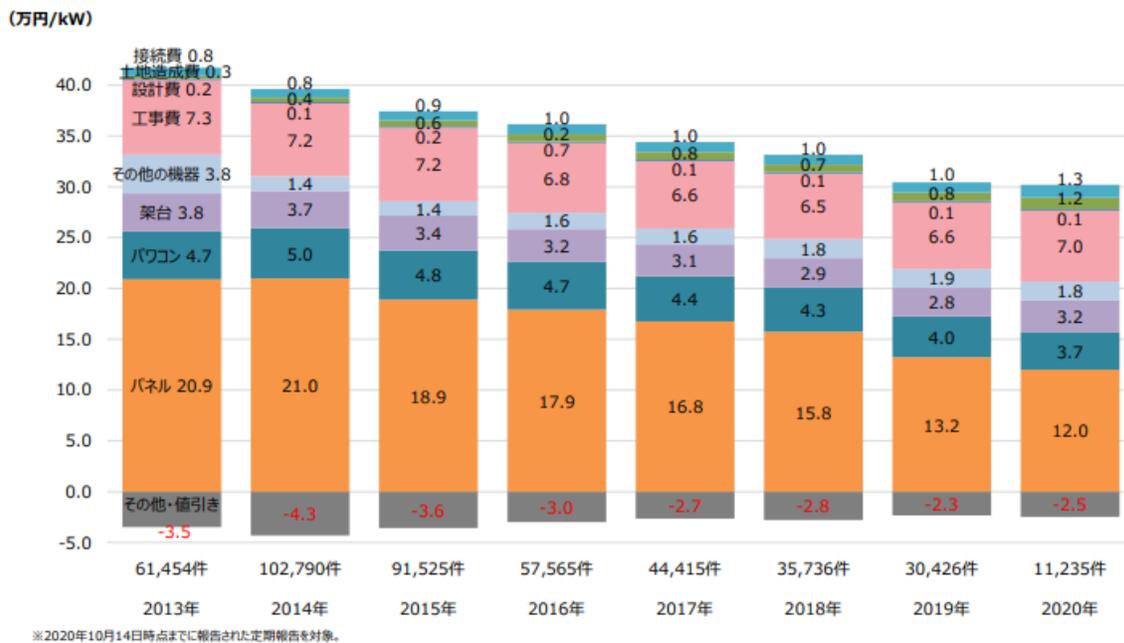

*Figure S10: Breakdown of capital costs for solar PV in Japan. Source: METI report.*

## Offshore wind

According to the [METI report](), average LCOE for offshore wind deployed in Japan during 2014-2019 was 36 yen/kWh (32.7 US cents/kWh). Starting from November 2020, fixed-bottom projects in the 4 offshore wind promotion zones are subjected to a maximum price of 29 yen/kWh (26.4 US cents/kWh). This is still significantly higher than international costs, which decreased from 23.3 yen/kWh (21.2 US cents/kWh) in 2014 to 8.6 yen/kWh (7.8 US cents/kWh) in 2020. Europe is leading the development of offshore wind, and large cost reductions are achieved along with the deployment of increasingly larger offshore wind turbines. Japan is expecting to reduce the LCOE for fixed bottom offshore wind to this level (8-9 yen/kWh or 7.3-8.2 US cents/kWh) by 2030-2035, according to the [Green Growth Strategy]() published by METI. However, there is limited shallow water within Japan's exclusive economic zone, and therefore majority of the future offshore wind turbines will be floating based. Although in Japan, the current feed-in-tariff (36 yen/kWh or 32.7 US cents/kWh) is the same for floating systems and fixed bottom systems, the former usually has higher capital costs due to the floating structure. Estimation of the 2020 costs for floating offshore wind in Japan is presented in [a study]() by Toshiki et al., and CAPEX is found to be 1,500 million yen/MW (US$13,636/kW) with an LCOE of 40 yen/kWh (36.3 US cents/kWh). However, as cited in [another REI report](), studies show that LCOE of floating offshore wind will be cut in half between 2019 and 2032, and by 2030 the LCOE would be similar to that for fixed bottom systems at below US$100/MWh (Figure S11). This is consistent with the findings from [a study done by Bosch et al.](), in which the average LCOE for the identified offshore wind resources in Japan is $86/MWh, with the lowest LCOE observed at deep water where floating foundations are used. In this study, we assume that massive deployment of floating offshore wind will lead to large cost reductions, and the LCOE for floating offshore wind will drop to the same level as fixed bottom systems at around 8 yen/kWh (7.3 US cents/kWh). Estimated future CAPEX and OPEX costs from [Toshiki et al.'s study]() are adopted in this study.

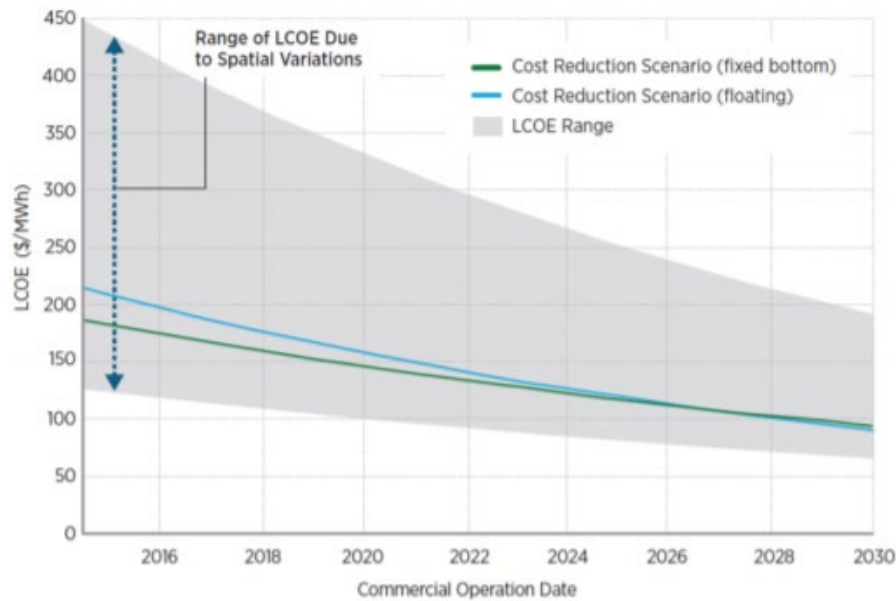

*Figure S11: Projections for LCOE for offshore wind. Source: REI report.*

## Pumped hydro energy storage

Costs for pumped hydro energy storage are adopted from a study by Stocks et al. Cost assumptions for Class A sites are used. In Japan, 3,295 GWh of the total 52,657 GWh pumped hydro capacity identified are Class A sites.

## Transmission

Costs for transmission, including HVDC (overhead and submarine) and HVAC, are adopted from a study by Lu et al. An exchange rate of 0.7 is used to convert AUD to USD.

## Nuclear, hydro and other renewables

Costs for nuclear, hydro and biomass are adopted from a report published by the Institute of Energy Economics (IEE) Japan, in which electricity generation costs in Japan for different technologies are calculated for 2014 and projected for 2030. LCOE for nuclear is expected to increase slightly from 10.1 yen/kWh (9.2 US cents/kWh) to 10.3 yen/kWh (9.4 US cents/kWh) in 2030. LCOE for hydro is expected to remain stable at 11 yen/kWh (10 US cents/kWh). LCOE for other renewables are higher (16.8 yen/kWh for geothermal and 29.7 yen/kWh for biomass) but will have a much lower impact due to the low installed capacity (4,427 MW combined).

## Hydrogen

Cost for hydrogen is adopted from [a study](#) by Longden et al., in which future cost of green hydrogen is estimated to be US$2-3/kg. This is consistent with the goal set in the [Green Growth Strategy](#), in which cost of hydrogen is expected to be below 30 yen/Nm$^3$ (US$3/kg) by 2030 and below 20 yen/ Nm$^3$ (US$2/kg) by 2050. Both values are modelled in this study, corresponding to US$108/MWh and US$163/MWh respectively, using 0.05427 as the kgH2/kWh conversion efficiency as quoted in [Australia's National Hydrogen Roadmap](#). It is also assumed that imported hydrogen will be converted to electricity in Japan via gas peaker, and cost assumptions (average of 'low' and 'high' cases) in [Lazard's report](#) are adopted, with the fuel price substituted by import price of hydrogen.

## Real discount rate

The following values are used in various studies for real discount rate in Japan:

*Table S4: Discount rates used in literature*

| Source | Value | Technology |
| --- | --- | --- |
| [Lazard's LCOE report](#) | 3.72% - 4.2% | All technologies |
| [IEE report](#) | 3% | All technologies |
| [REI solar report](#) | 3.2% | Solar PV |
| [Carbon tracker report](#) | 3.5% for solar PV and onshore wind<br>4.2% for offshore wind | Solar PV and wind |

In this study a real discount rate of 3.5% is used for all technologies.

## Supplementary Information D: Sensitivity analysis

For each scenario, the 'worst year' is determined by the energy deficit calculated from the optimized capacities (PV, wind, and storage), assuming only solar PV and wind are supplying the electricity demand. In another word, deficit is the amount of energy that would have been supplied by dispatchable sources (hydro and hydrogen). In the 'Simulation' module, hydro is set to run-off-river and hydrogen is set to zero. Moving average with a window of 168 hours (1 week) is calculated for the deficit. It represents the overall 'stress' on the system over any consecutive periods due to variable generation and demand. Results for the 'Baseline' scenario and the 'Wind-dominated' scenario are shown below.

The 'worst year' is defined as the year in which largest deficit moving average occurs, which is 1993 for the 'Baseline' scenario and 2006 for the 'Wind-dominated' scenario. Modelling the worst year only yields similar results compared with the full 40-year modelling for both the 'Baseline' scenario (both $85/MWh) and the 'Wind-dominated' scenario ($110/MWh vs $106/MWh).

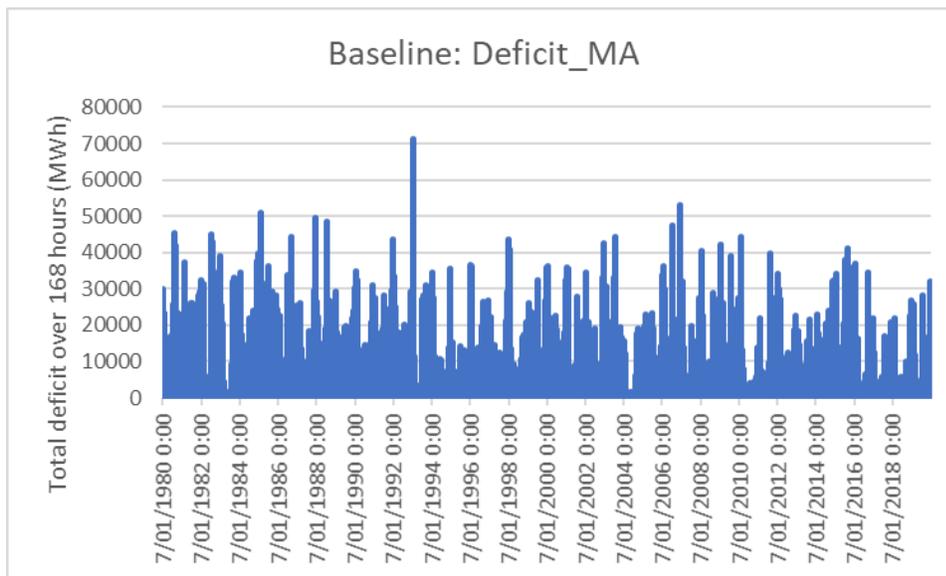

Figure S12: Baseline deficit moving average

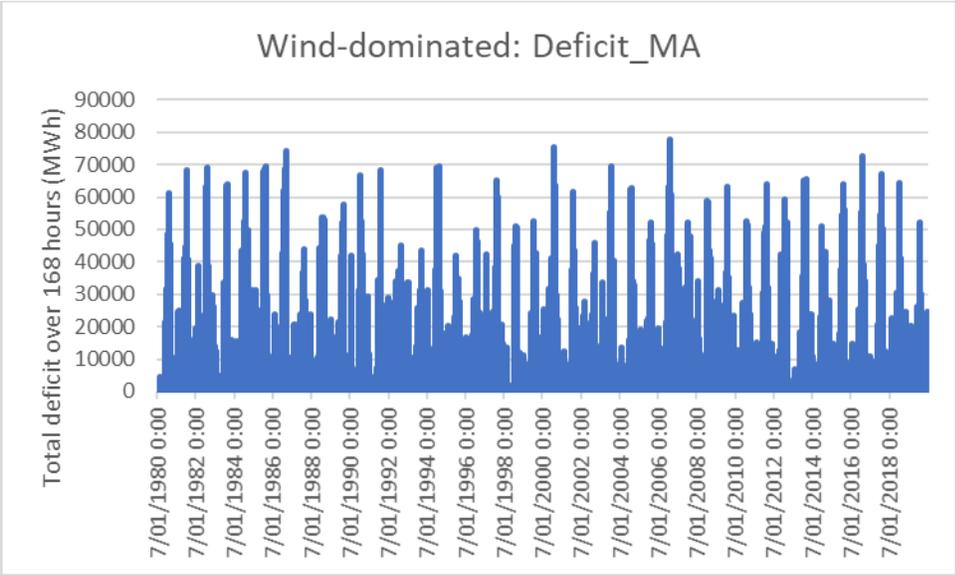

*Figure S13: Wind-dominated deficit moving average*